\documentclass{article}

\usepackage{arxiv}
\usepackage[utf8]{inputenc} 
\usepackage[T1]{fontenc}    
\usepackage{hyperref}       
\usepackage{url}            
\usepackage{booktabs}       
\usepackage{amsfonts}       
\usepackage{nicefrac}       
\usepackage{microtype}      
\usepackage{lipsum}		    
\usepackage{graphicx}
\usepackage{doi}
\usepackage{xcolor}
\usepackage{siunitx}
\usepackage{amsmath}
\usepackage{amssymb}
\usepackage{wrapfig}

\newcommand{\dalle}{DALL$\cdot$E-2~}
\hypersetup{
    colorlinks,
    linkcolor={red!50!black},
    citecolor={blue!50!black},
    urlcolor={blue!80!black}
}

\title{Perspective (In)consistency of Paint by Text}

\author{Hany Farid \\
	Department of Electrical Engineering and Computer Sciences \\
	School of Information \\
	University of California, Berkeley \\
	\texttt{hfarid@berkeley.edu}\\
}

\date{}

\begin{document}
\maketitle
\large

\begin{abstract}
	Type ``a sea otter with a pearl earring by Johannes Vermeer'' or ``a photo of a teddy bear on a skateboard in Times Square'' into OpenAI's \dalle paint-by-text synthesis engine and you will not be disappointed by the delightful and eerily pertinent results. The ability to synthesize highly realistic images -- with seemingly no limitation other than our imagination -- is sure to yield many exciting and creative applications. These images are also likely to pose new challenges to the photo-forensic community. Motivated by the fact that paint by text is not based on explicit geometric modeling, and the human visual system's often obliviousness to even glaring geometric inconsistencies, we provide an initial exploration of the perspective consistency of \dalle synthesized images to determine if geometric-based forensic analyses will prove fruitful in detecting this new breed of synthetic media.
\end{abstract}


\section{Introduction}
\label{sec:introduction}

Early examples of learning-based image synthesis (e.g.,~for example, ProGAN~\cite{karras2017progressive} and StyleGAN~\cite{karras2019style}) were able to synthesize -- in the absence of explicit 3-D models -- images of faces or other object categories (bedrooms, landscapes, etc.). While able to synthesize highly-realistic images (particularly later incarnations of StyleGAN~\cite{karras2020analyzing,karras2021alias}), the synthesis engines did not afford much control over the semantic contents of the synthesized image.

Paint by text~\cite{razavi2019generating,radford2021learning,bau2021paint,yu2022scaling}, on the other hand, affords seemingly limitless control of semantic content. OpenAI's \dalle\footnote{\url{https://openai.com/dall-e-2}}, for example, synthesizes images from text descriptions~\cite{aditya2022clip} (see also Google's Imagen\footnote{\url{https://imagen.research.google}} and Parti\footnote{\url{https://parti.research.google}}). Text descriptions ranging from ``an armchair in the shape of an avocado'' to ``an emoji of a baby penguin wearing a blue hat, red gloves, green shirt, and yellow pants,'' generate eerily pertinent images.

Because paint-by-text and other learning-based synthesis engines are not based on explicit modeling of the 3-D scene, lighting, or camera, it seems plausible that synthesized images may not adhere to the precise geometric scene properties found in natural or traditional computer-generated imagery (CGI). At the same time, such failures may not matter much to the human visual system which has been found to be surprisingly inept at certain geometric judgments including inconsistencies in lighting~\cite{ostrovsky05}, shadows~\cite{jacobson04,farid-bravo10}, reflections~\cite{bertamini03,farid-bravo10}, viewing position~\cite{banks05}, and perspective distortion~\cite{bravo-farid01,farid-bravo10}. 

With this as motivation, we provide an initial exploration of the perspective consistency of \linebreak \dalle synthesized images to determine if 3-D structures, cast shadows, and reflections are physically consistent with the expected perspective geometry of the 3-D to 2-D image formation process. 

This analysis reveals that while \dalle exhibits some basic understanding of perspective geometry, synthesized images contain consistent geometric inconsistencies which, while not always visually obvious, should prove forensically useful. It remains to be seen, however, if the next generation of synthesis engines will be able to fully learn the detailed geometry and physics of image formation.

\section{Photo Forensics: Perspective Analysis}
\label{sec:geometricforensics}

We begin by providing an overview of three related physics-based forensic analyses, each of which leverages the same underlying perspective geometry inherent to the image formation process (see~\cite{farid2016photo} for a more detailed exposition).

\subsection{Vanishing Points}
\label{subsec:vanishing-points}

\begin{figure}[t]
    \begin{center}
    \begin{tabular}{cc}
        \fbox{\includegraphics[width=0.4\linewidth]{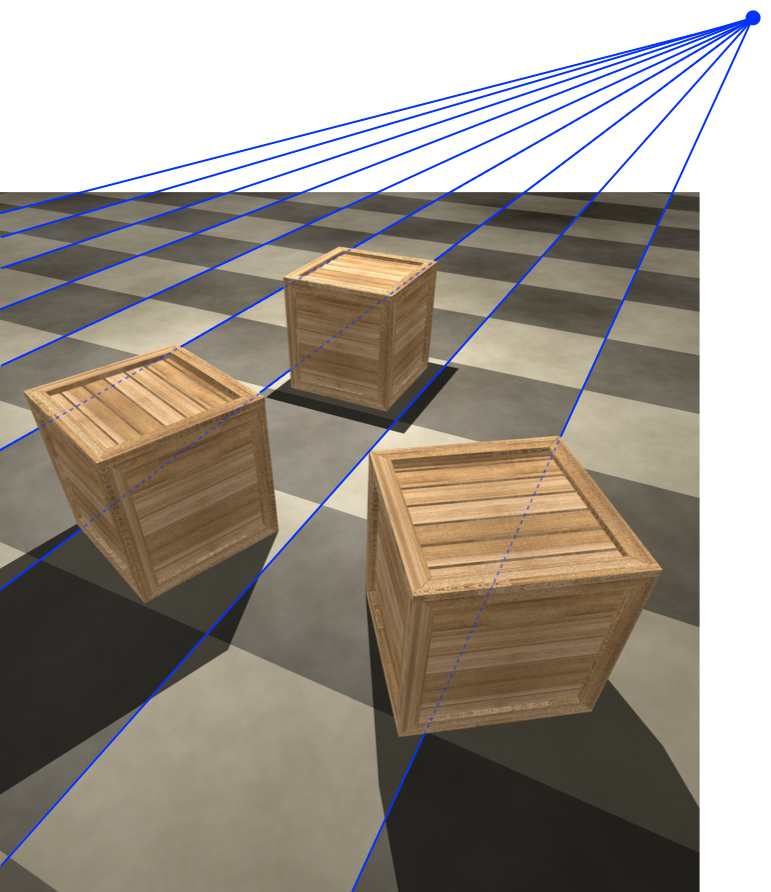}} &
        \fbox{\includegraphics[width=0.4\linewidth]{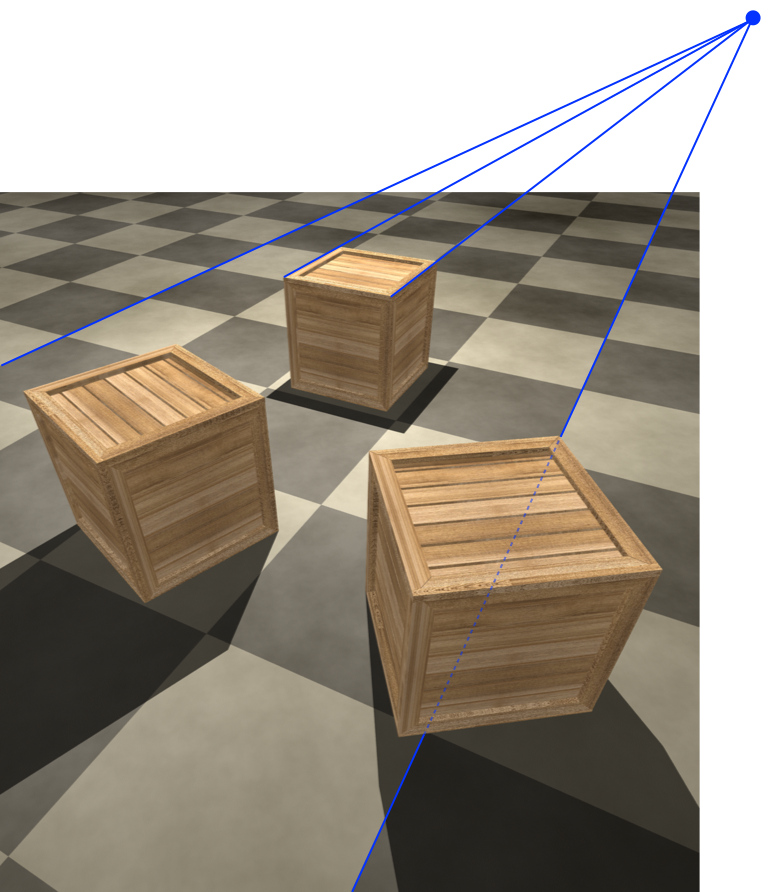}} \\
        (a) & (b) \\ \\
        \multicolumn{2}{c}{\fbox{\includegraphics[width=0.825\linewidth]{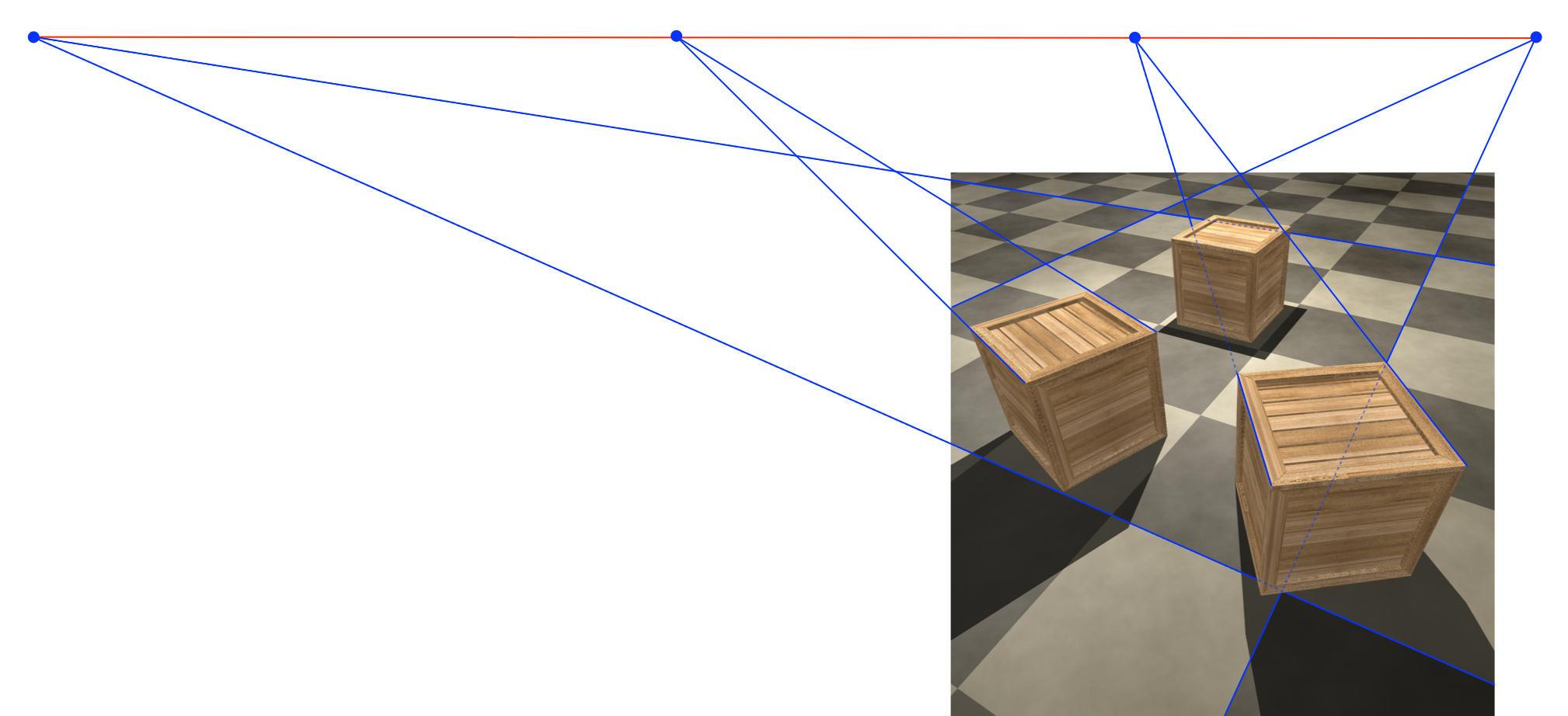}}} \\
        \multicolumn{2}{c}{(c)} \\
    \end{tabular}
    \end{center}
    \caption{Vanishing points and lines: (a) parallel lines on the same plane share a vanishing point; (b) parallel lines on parallel planes share a vanishing point; and (c) pairs of vanishing points (blue) on the same and parallel planes define a vanishing line (red).}
    \label{fig:vanishingpoints-constraint}
\end{figure}

If you imagine railroad tracks receding into the distance, you will likely visualize them converging to a single point -- a vanishing point -- on the horizon. We will consider four properties of the 3-D to 2-D projection of straight and parallel structures (see~\cite{hartley2003multiple} for a thorough coverage of projective and multi-view geometry):
\begin{enumerate}
    \item {\em Parallel receding lines converge at a vanishing point.} The lines between the tiles in Figure~\ref{fig:vanishingpoints-constraint}(a) are parallel in the scene. When imaged, these lines all converge at a vanishing point. If parallel lines in the scene recede in depth from the camera, then a vanishing point will exist, although it may fall outside the image. If parallel lines in the scene do not recede in depth—that is, if they are exactly parallel to the camera sensor (at any distance)—then the parallel lines are imaged as parallel lines, and for practical purposes we can consider the vanishing point to be at infinity. This geometry emerges from the basics of perspective projection. Under perspective projection a point $(X, Y, Z)$ in the scene is imaged to a point $(fX/Z, fY/Z)$ where $f$ is the camera focal length. Because the location of the point in the image is inversely proportional to the distance $Z$, the projected points compress as a function of distance, leading to converging lines in an image; 
    \item {\em Parallel lines on parallel planes converge to the same vanishing point.} The far box in Figure~\ref{fig:vanishingpoints-constraint}(b) is aligned with the tiles on the floor such that the edges of the box are parallel to the lines between the tiles. Because parallel lines on parallel planes share a vanishing point, the vanishing point for the sides of the box and the tile floor are the same; 
    \item {\em The vanishing points for all lines on a plane lie on a vanishing line.} An image may contain many sets of parallel lines, each converging to a different vanishing point, as shown in Figure~\ref{fig:vanishingpoints-constraint}(c). If the sets of parallel lines span the same plane in the scene, their vanishing points will lie on a vanishing line. The orientation of the vanishing line is determined by the rotation of the camera relative to the plane spanned by the parallel lines; and 
    \item {\em The vanishing points for all lines on parallel planes lie on a vanishing line.} The vanishing point for sets of parallel lines on parallel planes will lie on the same vanishing line. Because the tops of the boxes are parallel to the tiled floor, the vanishing points for each of the boxes in Figure~\ref{fig:vanishingpoints-constraint}(c) lie on the vanishing line defined by the vanishing points of the tiled floor.
\end{enumerate}

The plausibility of a vanishing point estimated from three or more lines (or a vanishing line estimated from three or more vanishing points) can be determined from the consistency of the constraints with a single solution. Small deviations from a consistent solution may reflect inaccuracies in the specification of the lines or small amounts of lens distortion, while large deviations likely indicate a physical inconsistency.

Each of the geometric constraints described above involves calculating vanishing points calculated from two or more lines. Two approaches for computing a vanishing point are described next, one for the simple case of two lines and one that involves estimation from three or more lines (see ~\cite{farid2016photo} for a more detailed exposition). 

In the simplest case, a vanishing point is computed from the intersection of two lines. The first line $l_1$ can be specified using a parametric equation: $l_1(t) = (\vec{p_1} - \vec{q_1})t + \vec{q_1}$, where $\vec{p_1} = (x_1,y_1)$ and $\vec{q_1} = (u_1,v_1)$ are any two distinct points on the line. The line connecting these two points is constructed by varying the parametric variable $t$ between $-\infty$ and $\infty$. The second line is defined similarly as $l_2(t) = (\vec{p_2} - \vec{q_2})t + \vec{q_2}$, where $\vec{p_2} = (x_2,y_2)$ and $\vec{q_2} = (u_2,v_2)$ are any two distinct points on the second line.

The intersection of these two lines is determined by finding the parametric variables $t_1$ and $t_2$ for which the two lines evaluate to the same point, yielding the solution:
\begin{eqnarray}
\begin{pmatrix} t_1 \cr t_2 \end{pmatrix} & = &
\begin{pmatrix} x_1-u_1 & -(x_2-u_2) \cr y_1-v_1 & -(y_2-v_2) \end{pmatrix}^{-1}
\begin{pmatrix} u_2 - u_1 \cr v_2 - v_1 \end{pmatrix}.
\label{eqn:geometric-intersection-of-two-lines}
\end{eqnarray}
The intersection of the two lines is then specified by either $l_1(t_1)$  or  $l_2(t_2)$. If the lines $l_1$ and $l_2$ are not parallel, then the $2 \times 2$ matrix on the right-hand side of the above solution will be invertible. If the lines are parallel, then the lines do not have a finite intersection and the matrix will not be invertible. 

The accuracy of the vanishing point calculation may be increased by using more than two lines. While two non-parallel lines must intersect, three or more non-parallel lines may not have a single intersection. In such situations we seek the point that is closest to all lines. We will use the perpendicular distance between the point and each line, as opposed to the vertical or horizontal distance, so that the estimation is not dependent on the orientation of the lines. 

The perpendicular distance between a point $\vec{v}$ and a line $l$ defined by the points $\vec{p}$ and $\vec{q}$ is \linebreak $d = \|\vec{n}^T (\vec{v} - \vec{p}) \|$, where $\| \cdot \|$ denotes vector norm and $\vec{n}$ is a unit vector perpendicular to this line. The point $\vec{v}$ that minimizes the average distance to $m$ lines is determined by minimizing the quadratic error function:
\begin{eqnarray}
    E(\vec{v}) & = & \sum_{i=1}^{m} \| \vec{n_i}^T (\vec{v} - \vec{p_i}) \|^2,
\end{eqnarray}
where the point $\vec{p_i}$ and vector $\vec{n_i}$ parameterize the $i^{th}$ line. The point $\vec{v}$ that minimizes the average distance to all $m$ lines is determined by differentiating this error function with respect to $\vec{v}$, setting equal to $\vec{0}$, and solving for $\vec{v}$ to yield:
\begin{eqnarray}
    \vec{v} & = & \left( \sum_{i=1}^{m} \vec{n_i} \vec{n_i}^T \right)^{-1} \sum_{i=1}^{m} \vec{n_i} \vec{n_i}^T  \vec{p_i}.
\end{eqnarray}
The $2 \times 2$ matrix in this solution will be invertible unless all $m$ lines are parallel.

As we will see in the next two sections, this basic geometry of intersecting lines extends to an analysis of cast shadows and mirrored reflections.

%
%
\begin{wrapfigure}{r}{6cm}
\vspace{-0.4cm}
\fbox{\includegraphics[width=6cm]{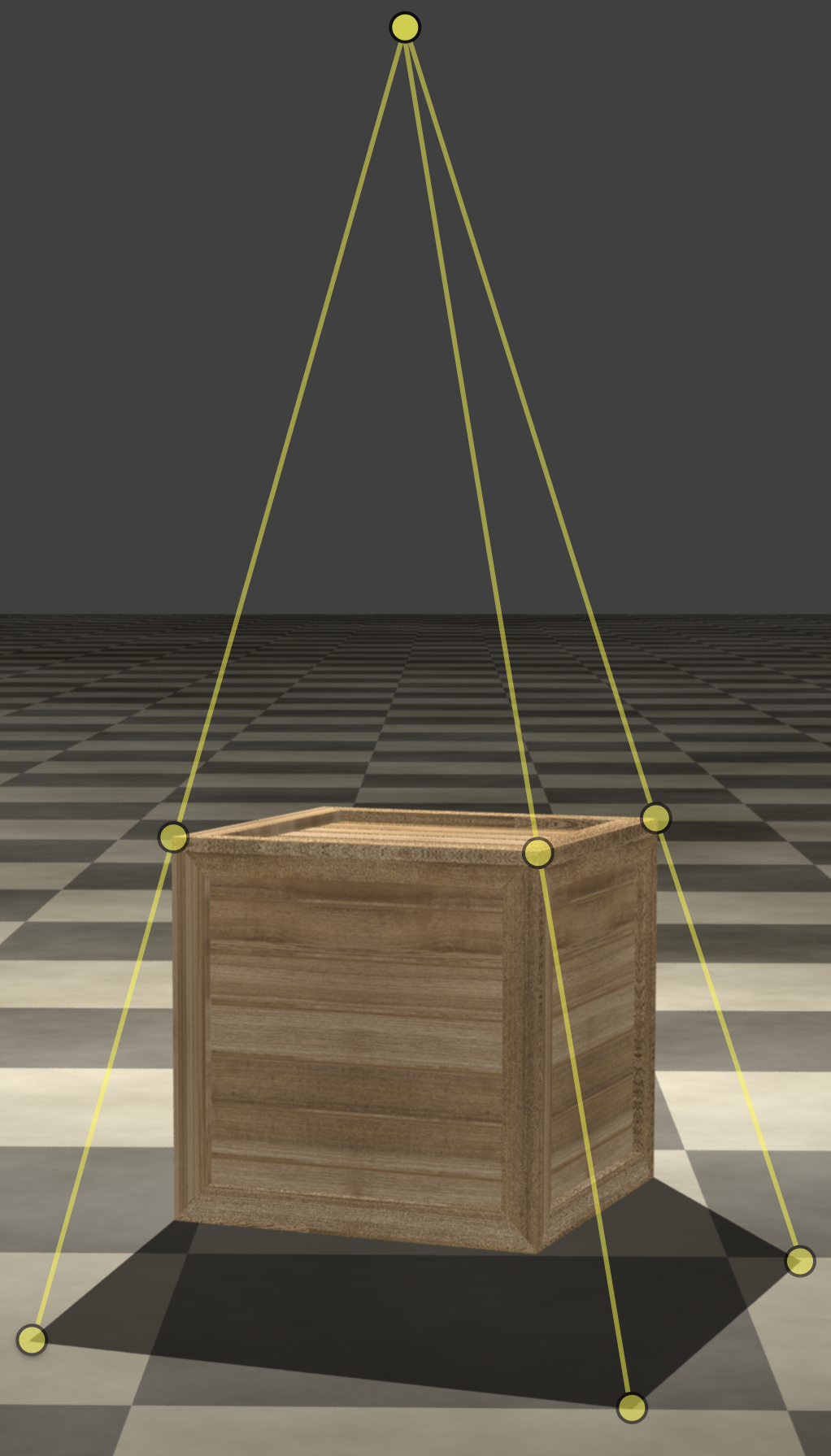}}
\end{wrapfigure}

\subsection{Shadows}
\label{subsec:shadows}

Somewhat surprisingly, the same geometry underlying vanishing points applies to cast shadows. Shown on the right are three rays connecting points on the box and their corresponding points on the cast shadow. Extending the image boundary reveals that these three rays intersect at a single point corresponding to the projection of the light source illuminating the scene. This geometric constraint relating the shadow, the object, and the light holds whether the light source is nearby (a desk lamp) or distant (the sun), and holds regardless of the location and orientation of the surfaces onto which shadows are cast. This analysis does, of course, assume a scene is illuminated by a single dominant light source, as would be evident from the presence of just a single cast shadow per object.

In the above example, the light source illuminating the scene is in front of the camera and so the projection of the light source is in the upper half of the image plane. If, however, the light is behind the camera then the projection of the light source will be in the lower half of the image plane. Because of this inversion, the shadow to object constraints must also be inverted. 

The cast shadow analysis of an image must, therefore, consider three possibilities: (1) the light is in front of the camera, the projection of the light source is in the upper half of the image plane, and the constraints are anchored on the cast shadow and encompass the object; (2) the light is behind the camera, the projection of the light source is in the lower half of the image plane, and the constraints are anchored on the object and encompass the cast shadow; or (3) the light is directly above or below the camera center, the projection of the light source is at infinity, and the constraints will intersect at infinity. If any of these cases lead to a common intersection of all constraints, then the cast shadows are physically plausible (see~\cite{kee2013exposing} for a more detailed exposition of this forensic technique).

\subsection{Reflections}
\label{subsec:reflections}

\begin{figure}[p]
    \begin{center}
    \begin{tabular}{c}
        \fbox{\includegraphics[width=0.8\linewidth]{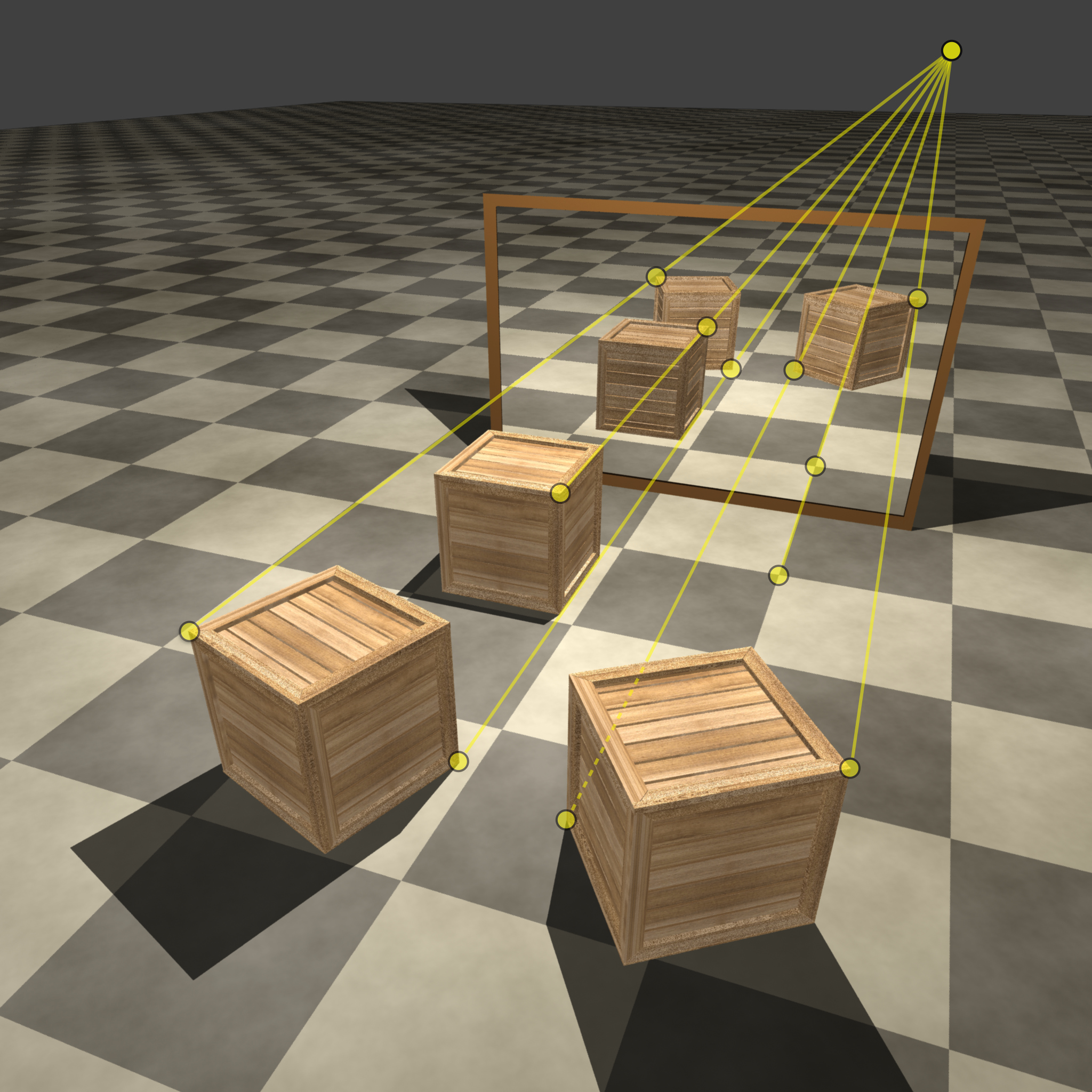}} \\
        \\
        \fbox{\includegraphics[width=0.8\linewidth]{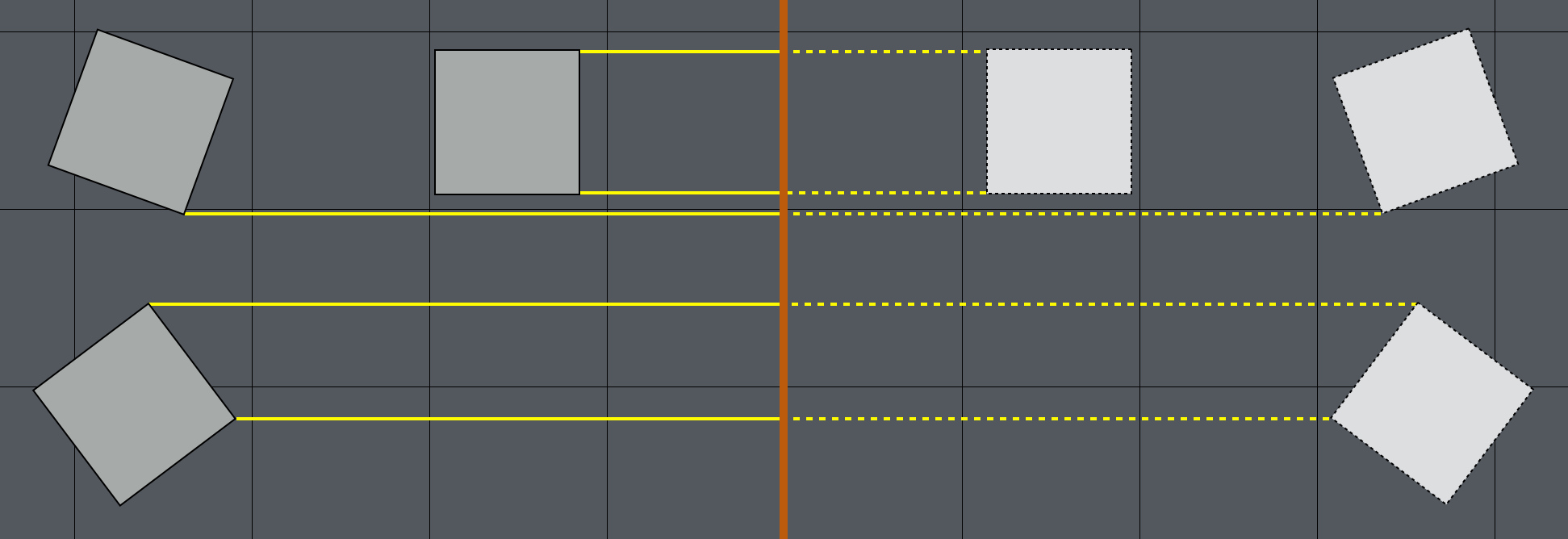}}    
    \end{tabular}
    \end{center}
    \caption{Three boxes are reflected in a mirror (top) and a virtual bird's eye view (bottom) of this scene with the boxes on the left, mirror in the middle, and reflection on the right. The yellow lines connect points on a box to their reflection.}
\label{fig:physics-reflections-boxes-1}
    \label{fig:reflections-constraint}
\end{figure}

Shown in Figure~\ref{fig:reflections-constraint} is a scene in which three boxes are reflected in a flat mirror. The lower portion of this figure shows the geometric relationship between the real boxes and the virtual boxes. The orange line represents the mirror, located at the midpoint between the two sets of boxes. The yellow lines connect corresponding points on the real and virtual boxes. These lines are parallel to each other and perpendicular to the mirror.  Now let's consider how these parallel lines appear when they are superimposed on the scene. The lines that were parallel when viewed from the plane of the mirror are no longer parallel. Instead, due to perspective projection, these parallel lines converge to a single point, just as parallel lines in the world converge to a vanishing point (see Section~\ref{subsec:vanishing-points}). Because the lines connecting corresponding points in a scene and their reflection are always parallel, these lines must have a common intersection in the image to be physically plausible (see~\cite{obrien2012exposing} for a more detailed exposition of this forensic technique).

\section{Paint by Text: Perspective Analysis}

A collection of images with varying text prompts were synthesized using OpenAI's \dalle (as compared to the less powerful DALL$\cdot$E Mini\footnote{\url{https://dallemini.com}}). The text prompts were selected so as to elicit images with geometric perspective, cast shadows, and mirrored reflections. This space of text prompts is, of course, nearly infinite, and we will only explore a tiny subset of this space. Although limited, the synthesis examples provided below do not appear to be an artifact of the specific text prompts: similar results were achieved with a range of different prompts.

\subsection{Vanishing Points}
\label{subsec:vanishing-points-analysis}

\begin{wrapfigure}{r}{6cm}
\vspace{-0.4cm}
\fbox{\includegraphics[width=6cm]{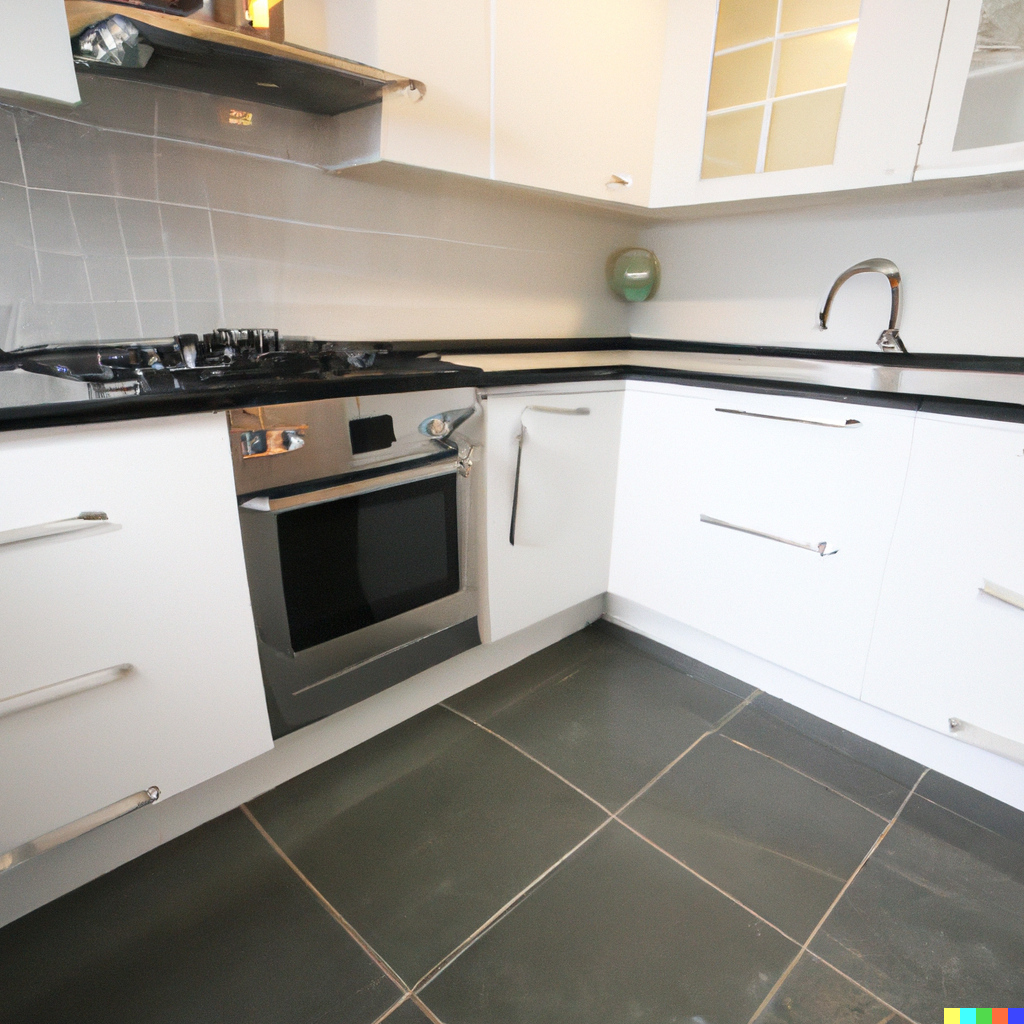}}
\end{wrapfigure}
Shown to the right is a representative example of an image synthesized with the text prompt ``A photo of a kitchen with a tiled floor.'' Although some of the details on the counter-top and cabinets are synthesized incorrectly, \dalle synthesizes the scene under what would appear, at first glance, to be appropriate perspective.  

\begin{figure}[t]
    \begin{center}
    \begin{tabular}{c@{\hspace{0.12cm}}c}
        \includegraphics[height=5.25cm]{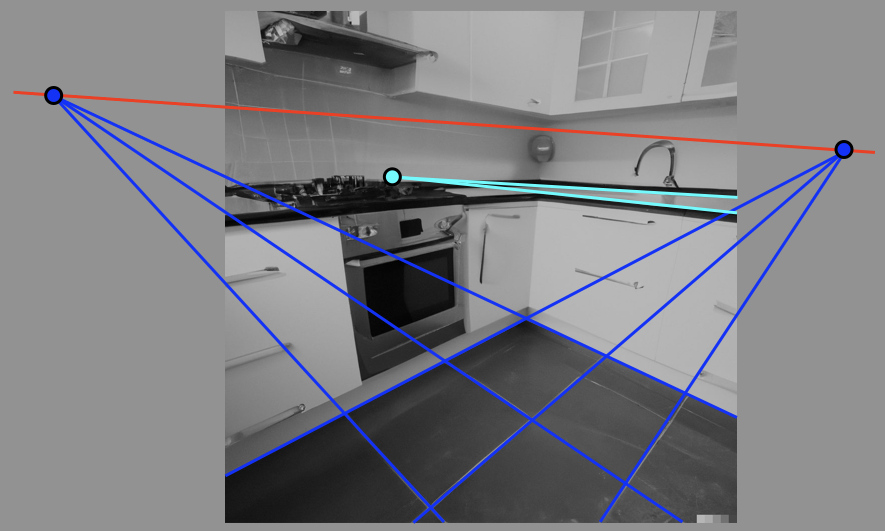} &
        \includegraphics[height=5.25cm]{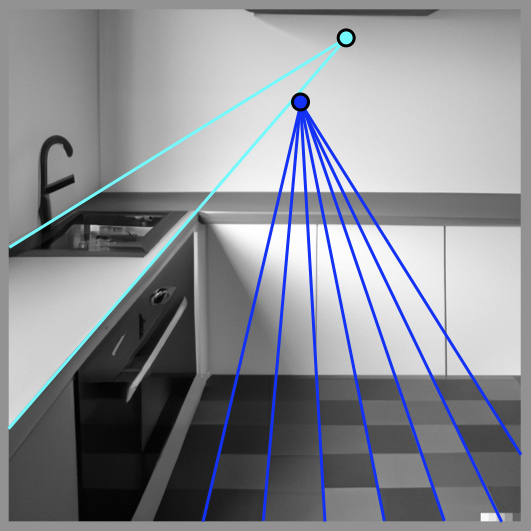} \\
        \multicolumn{2}{c}{\includegraphics[width=0.85\linewidth]{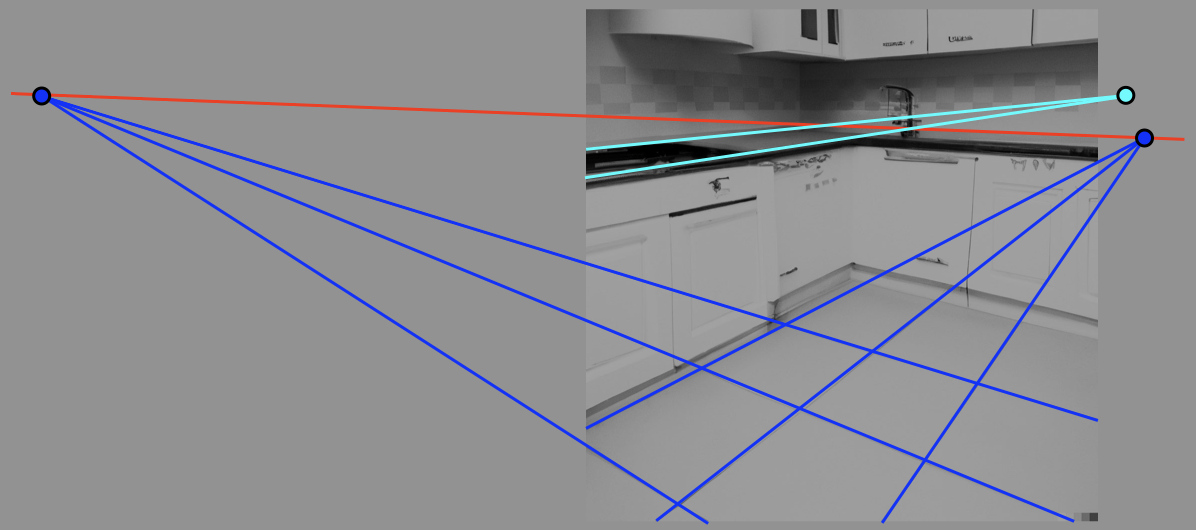}}
    \end{tabular}
    \end{center}
    \caption{Each set of parallel lines on the tiled floor are geometrically consistent, intersecting at a single vanishing point (blue). The vanishing point of the parallel counter-top (cyan), however, is inconsistent with both the vanishing line (red) and the vanishing point defined by the corresponding aligned tiles.}
    \label{fig:vanishingpoints-analysis}
\end{figure}

Shown in Figure~\ref{fig:vanishingpoints-analysis} are three representative examples of these images overlaid with an analysis of the consistency of the geometric perspective on the floor and counter top. Each image -- within a few pixel -- accurately captures the perspective geometry of the tiled floors, as evidence by the consistent vanishing points (rendered in blue). The vanishing point of the parallel counter-top (rendered in cyan), however, is geometrically inconsistent with the vanishing point of the corresponding aligned tiles (see Section~\ref{subsec:vanishing-points}). Even if the counter top were not parallel to the tiles, the cyan-colored vanishing point should lie on the vanishing line (rendered in red) defined by the tiled-floor's vanishing points. Note that for the image in the top right of Figure~\ref{fig:vanishingpoints-analysis}, the horizontal lines on the tiled floor are nearly parallel and so the corresponding vanishing point is at infinity and, therefore, not rendered.

While the perspective in these images is -- impressively -- locally consistent, it is not globally consistent. This same pattern was found in each of $25$ synthesized kitchen images.

%
%

\subsection{Shadows}
\label{subsec:shadows-analysis}

\begin{figure}[p]
    \begin{center}
    \begin{tabular}{c@{\hspace{0.12cm}}c@{\hspace{0.12cm}}c}
        \includegraphics[width=0.275\linewidth]{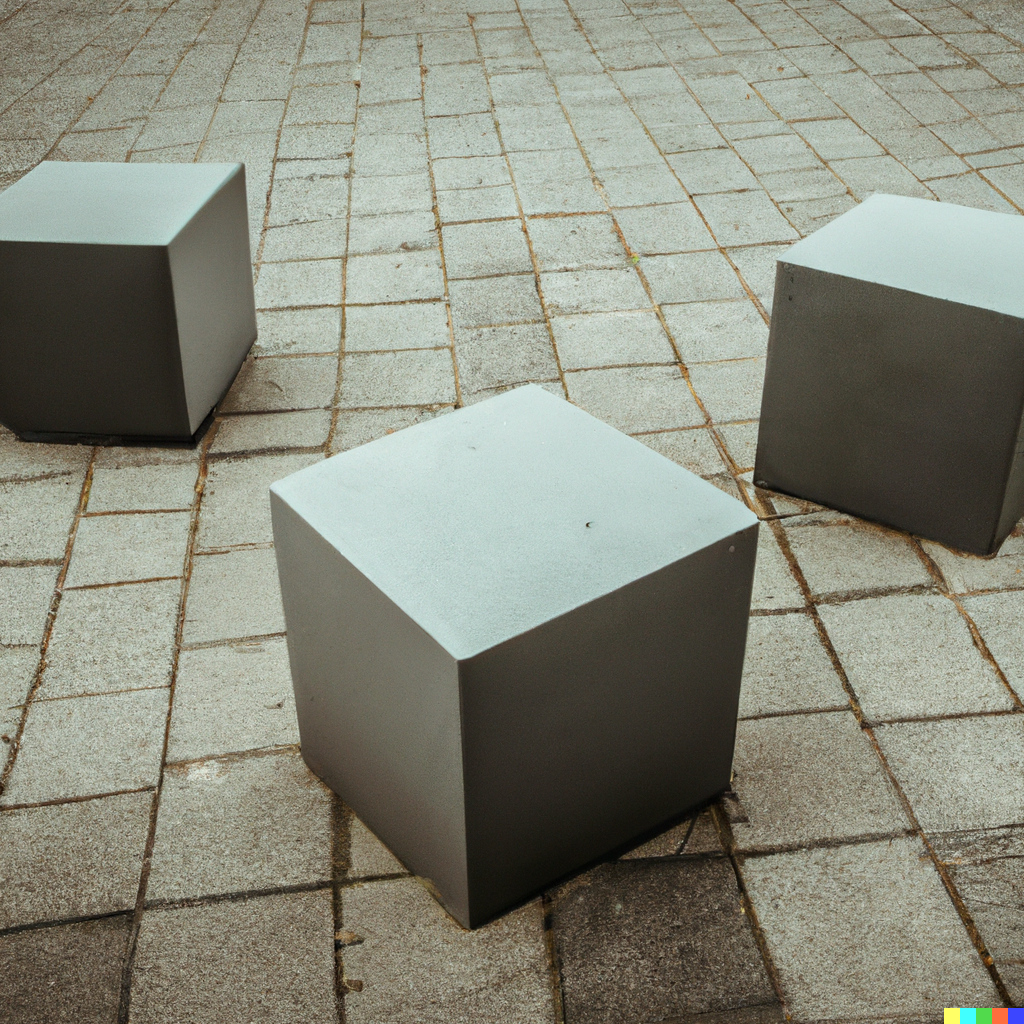} &
        \includegraphics[width=0.275\linewidth]{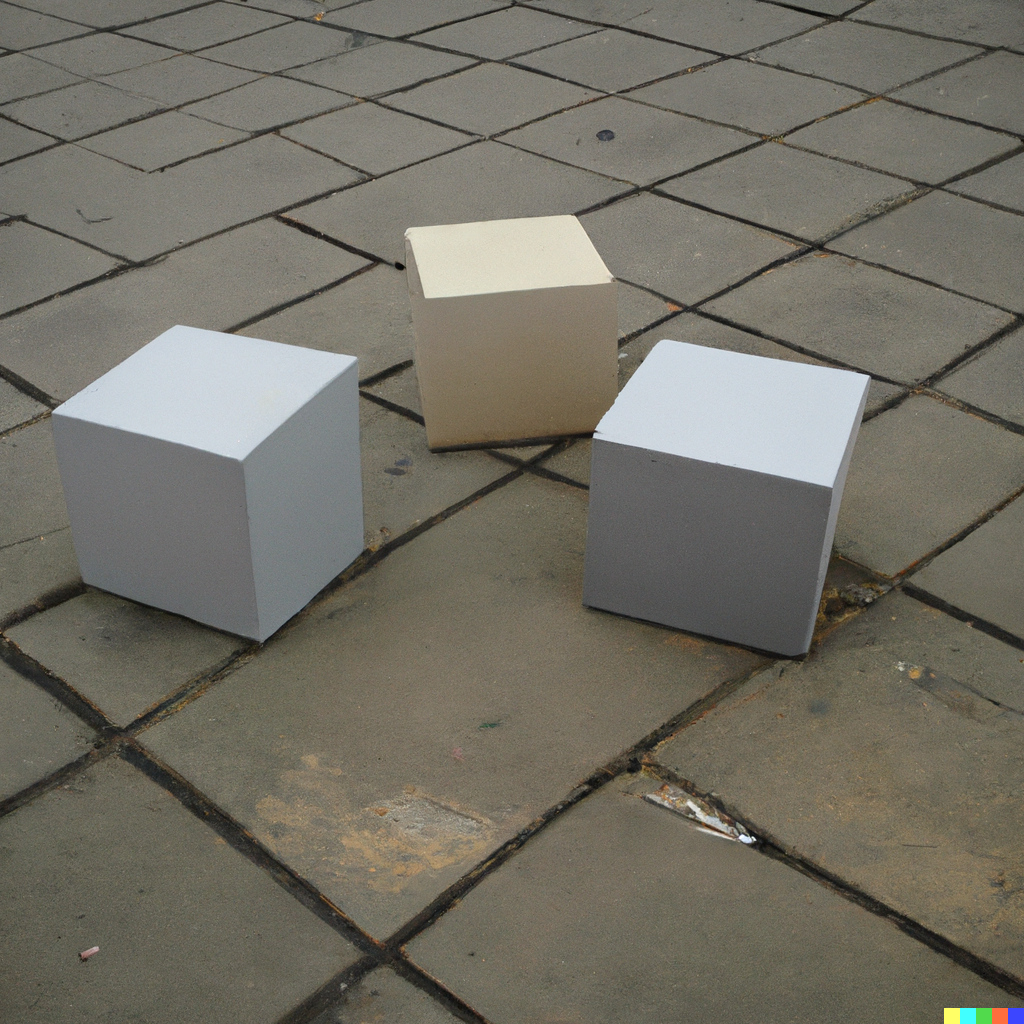} &
        \includegraphics[width=0.275\linewidth]{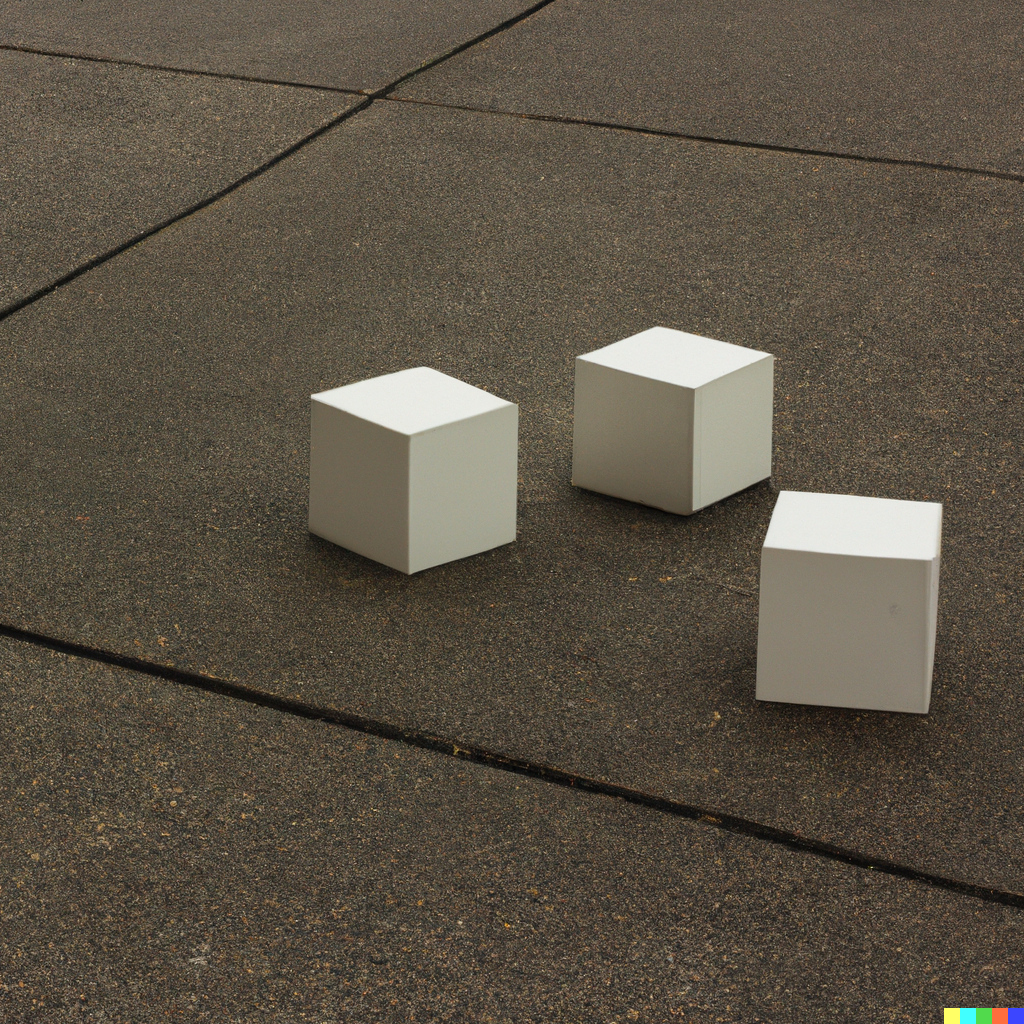} \\
        \includegraphics[width=0.275\linewidth]{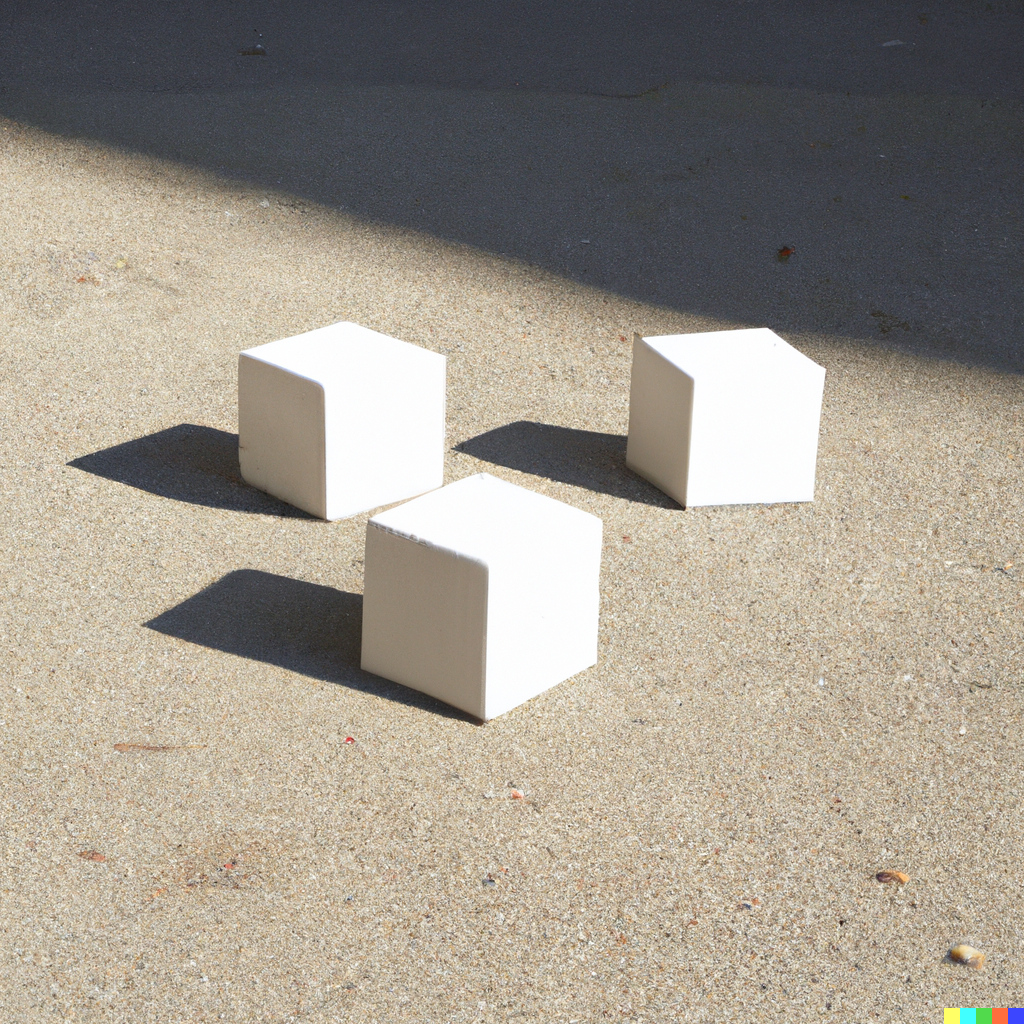} &
        \includegraphics[width=0.275\linewidth]{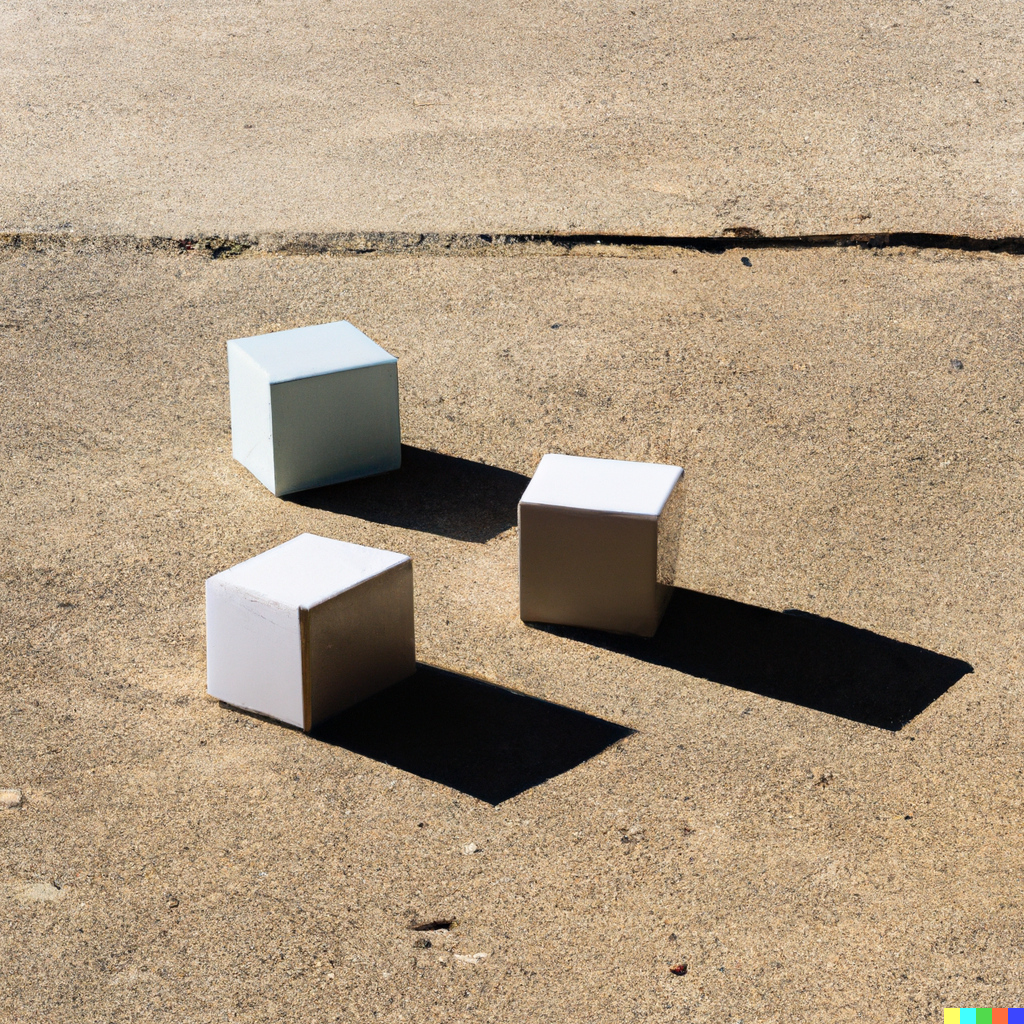} & 
        \includegraphics[width=0.275\linewidth]{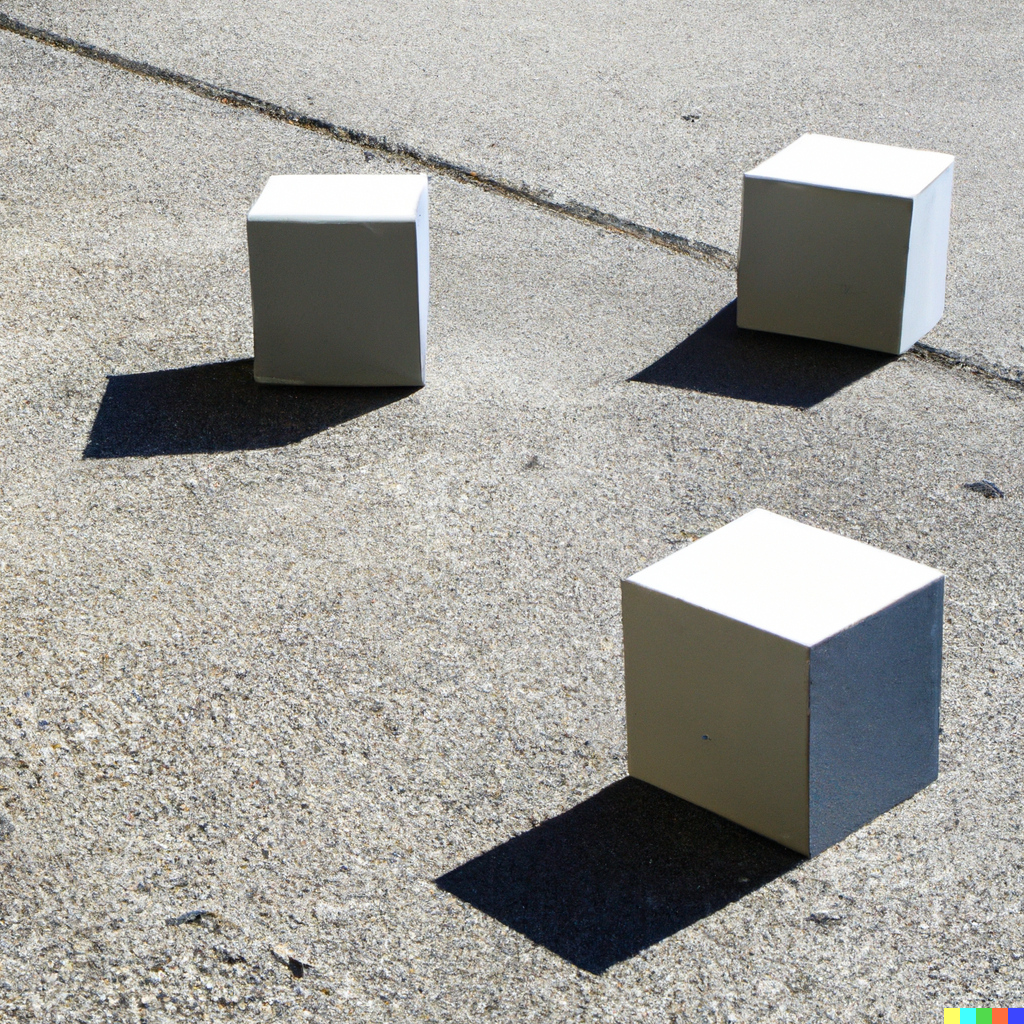} 
    \end{tabular}
    \end{center}
    \caption{Representative examples of synthesized images with the text prompt ``three cubes on a sidewalk photographed on a cloudy day'' (top) and ``three cubes on a sidewalk photographed on a sunny day'' (bottom).}
    \label{fig:shadows-cloudy-sunny}
\end{figure}

Shown in the top row of Figure~\ref{fig:shadows-cloudy-sunny} are three representative examples of synthesized images with the text prompt ``three cubes on a sidewalk photographed on a cloudy day.'' Shown in the bottom row of this figure are three representative examples of synthesized images with the text prompt ``three cubes on a sidewalk photographed on a sunny day.'' With no shadows in the cloudy-day images and crisp cast shadows in the sunny-day images, \dalle is capable of synthesizing images with appropriate generic lighting specifications. Further, at first glance, the cast shadows seem to be reasonably consistent with the object shape and a single light source (e.g.,~the sun).

\begin{figure}[p]
    \begin{center}
    \begin{tabular}{c@{\hspace{0.12cm}}c@{\hspace{0.12cm}}c}
        \includegraphics[width=0.275\linewidth]{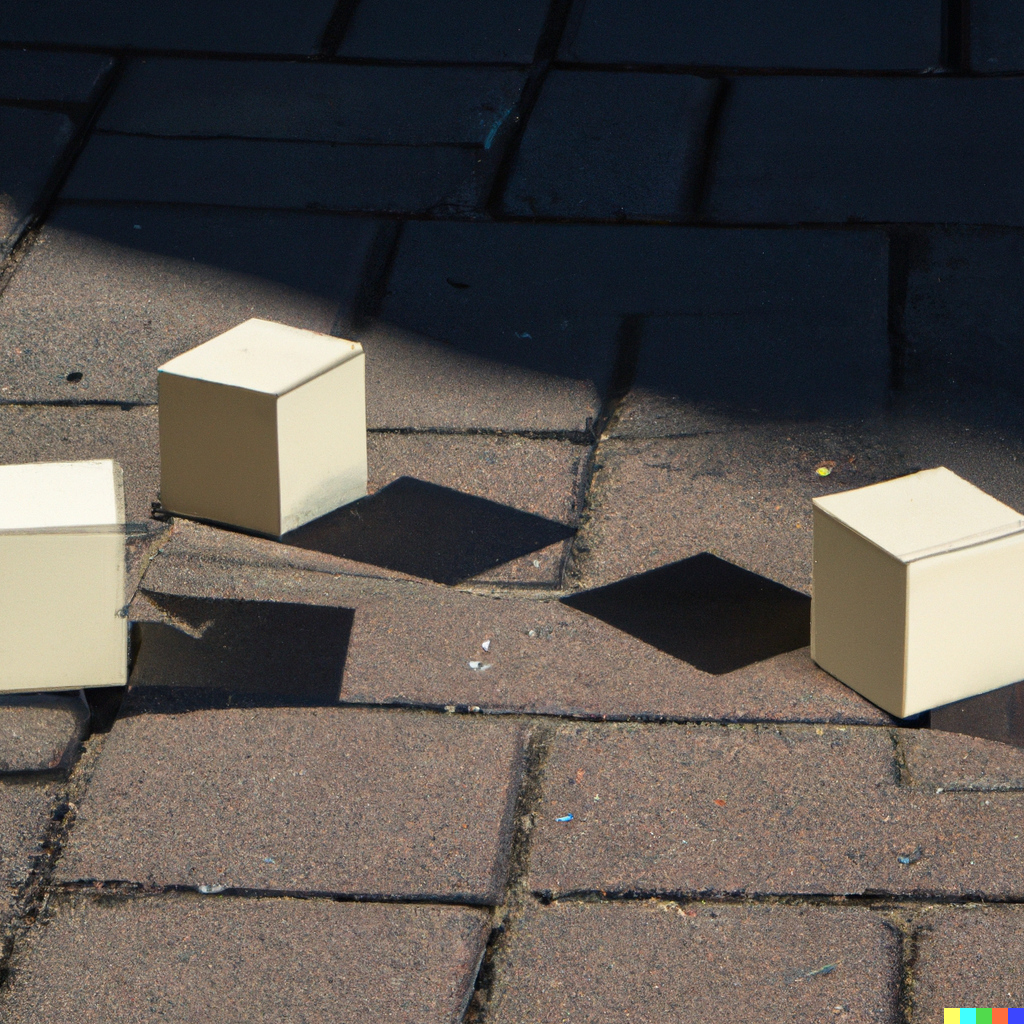} &
        \includegraphics[width=0.275\linewidth]{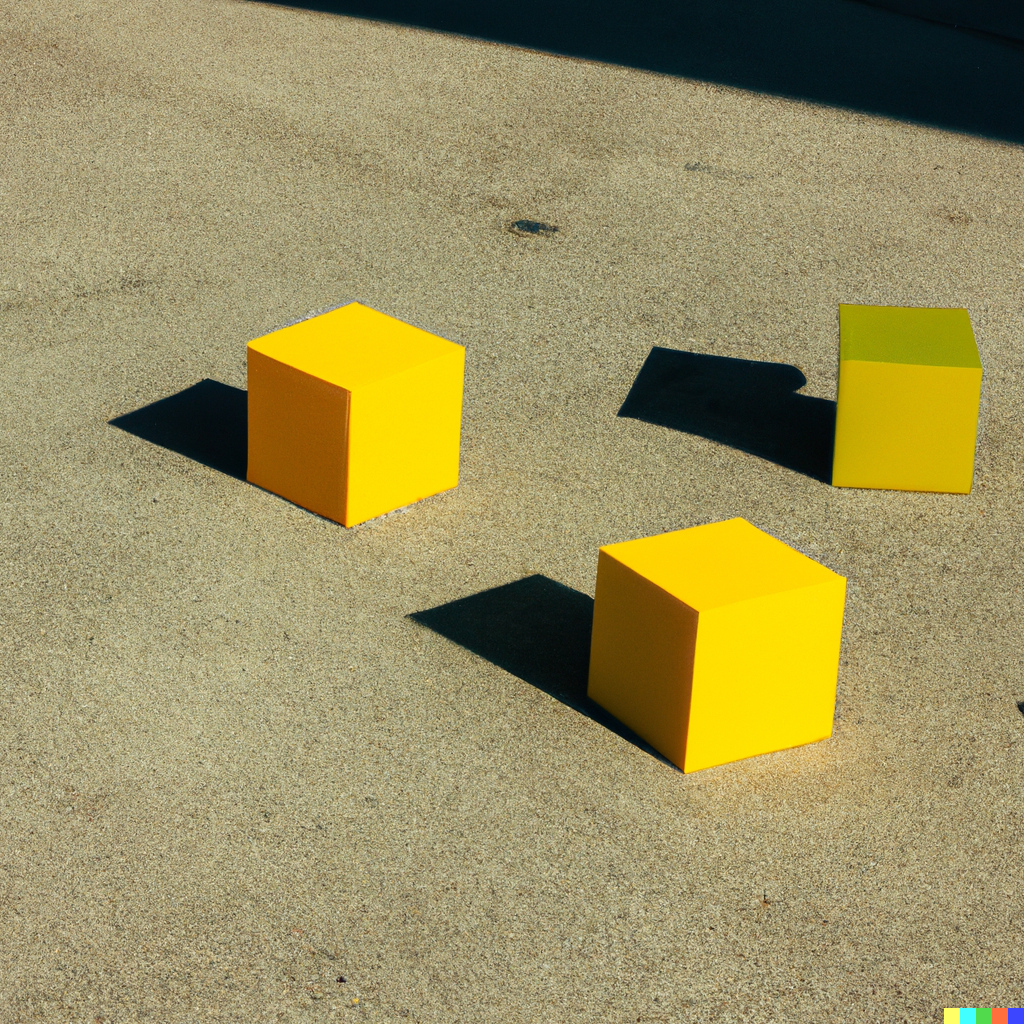} &
        \includegraphics[width=0.275\linewidth]{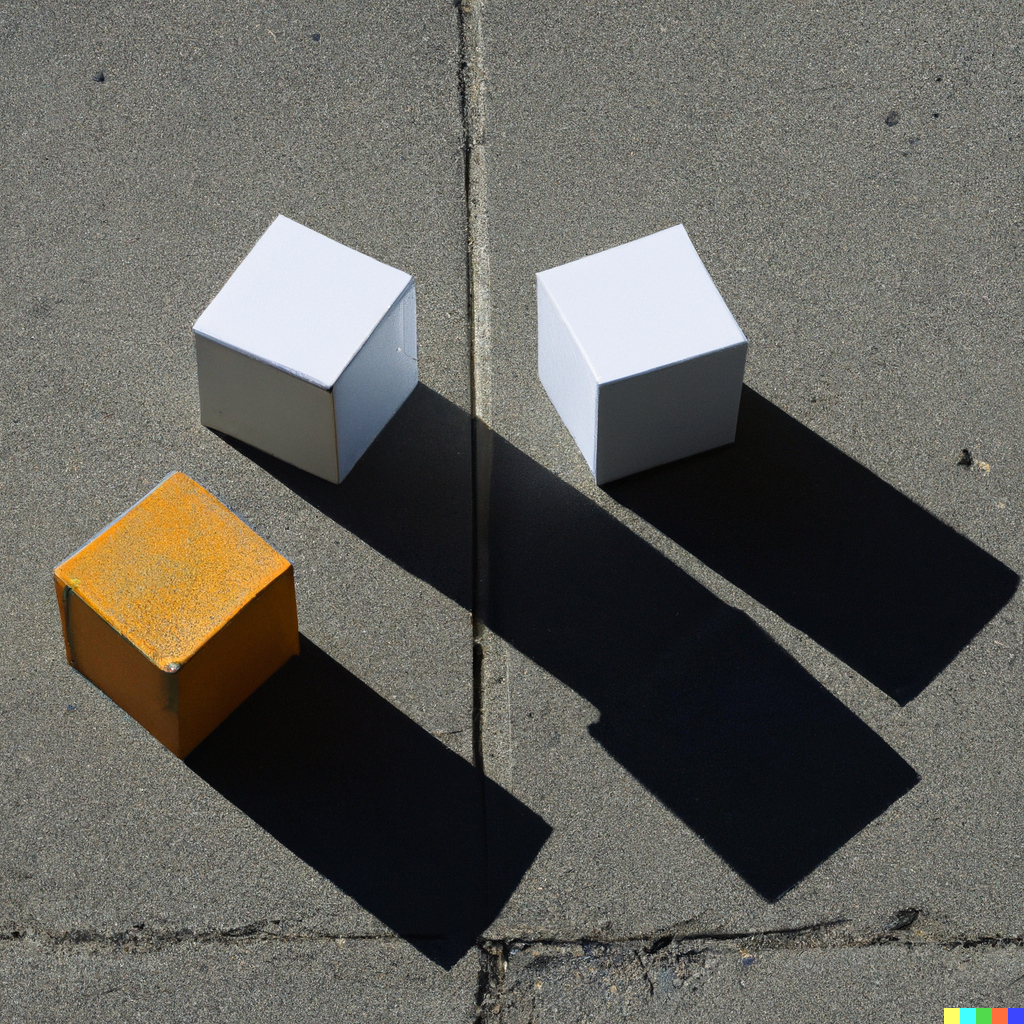} \\
        \includegraphics[width=0.275\linewidth]{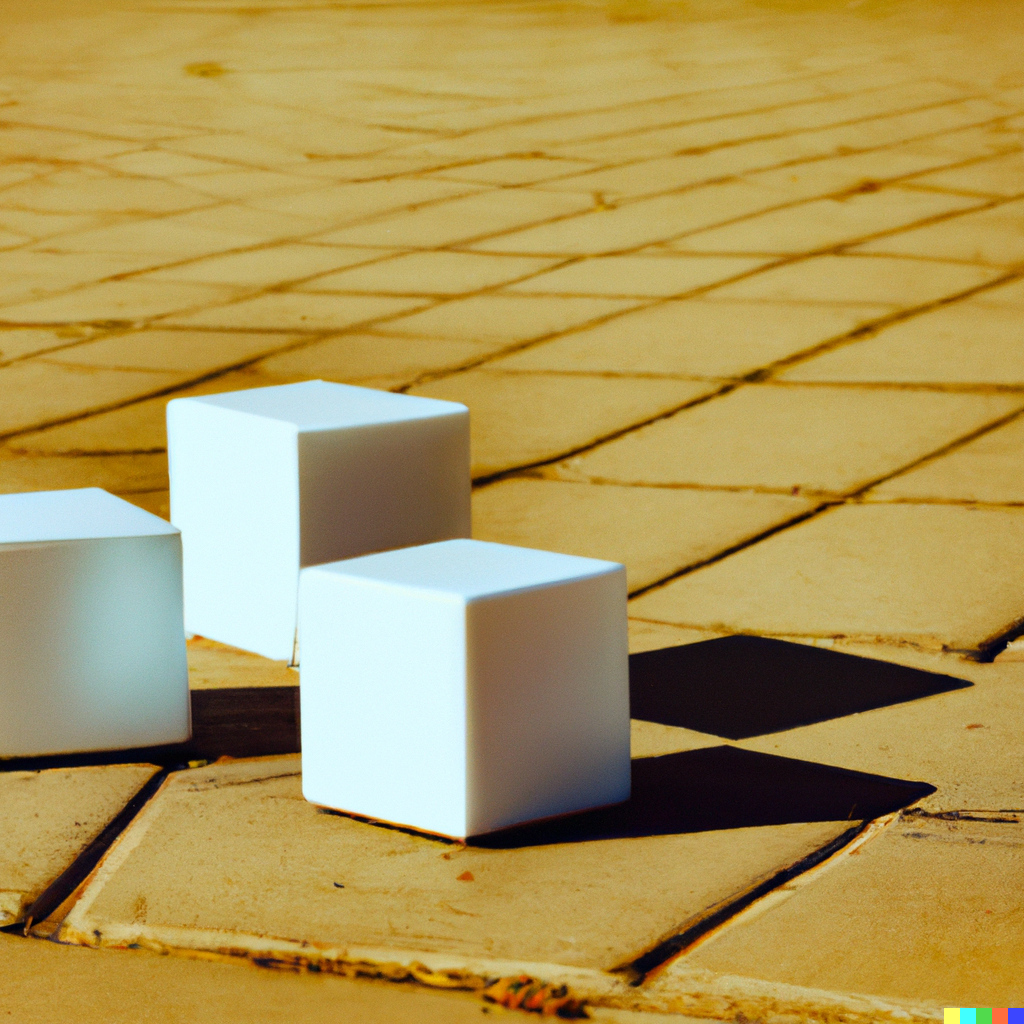} &
        \includegraphics[width=0.275\linewidth]{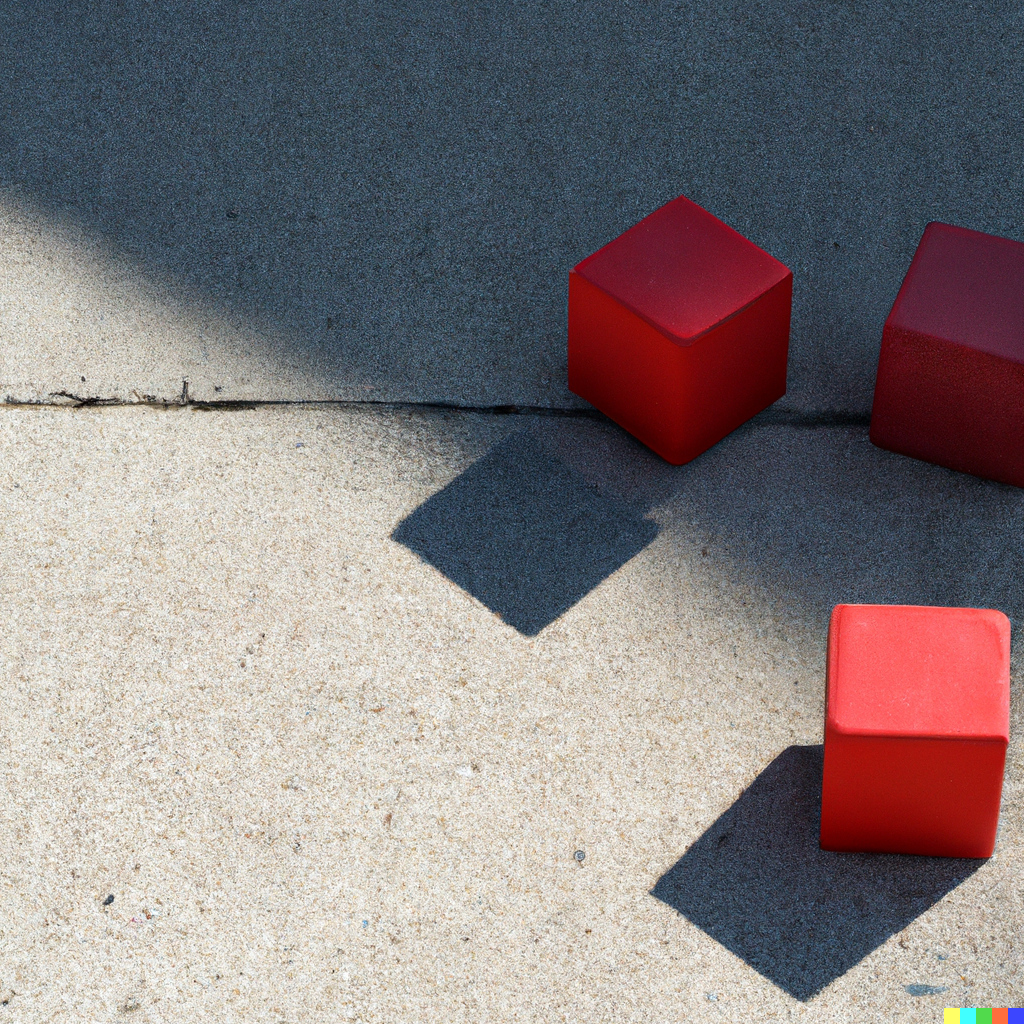} & 
        \includegraphics[width=0.275\linewidth]{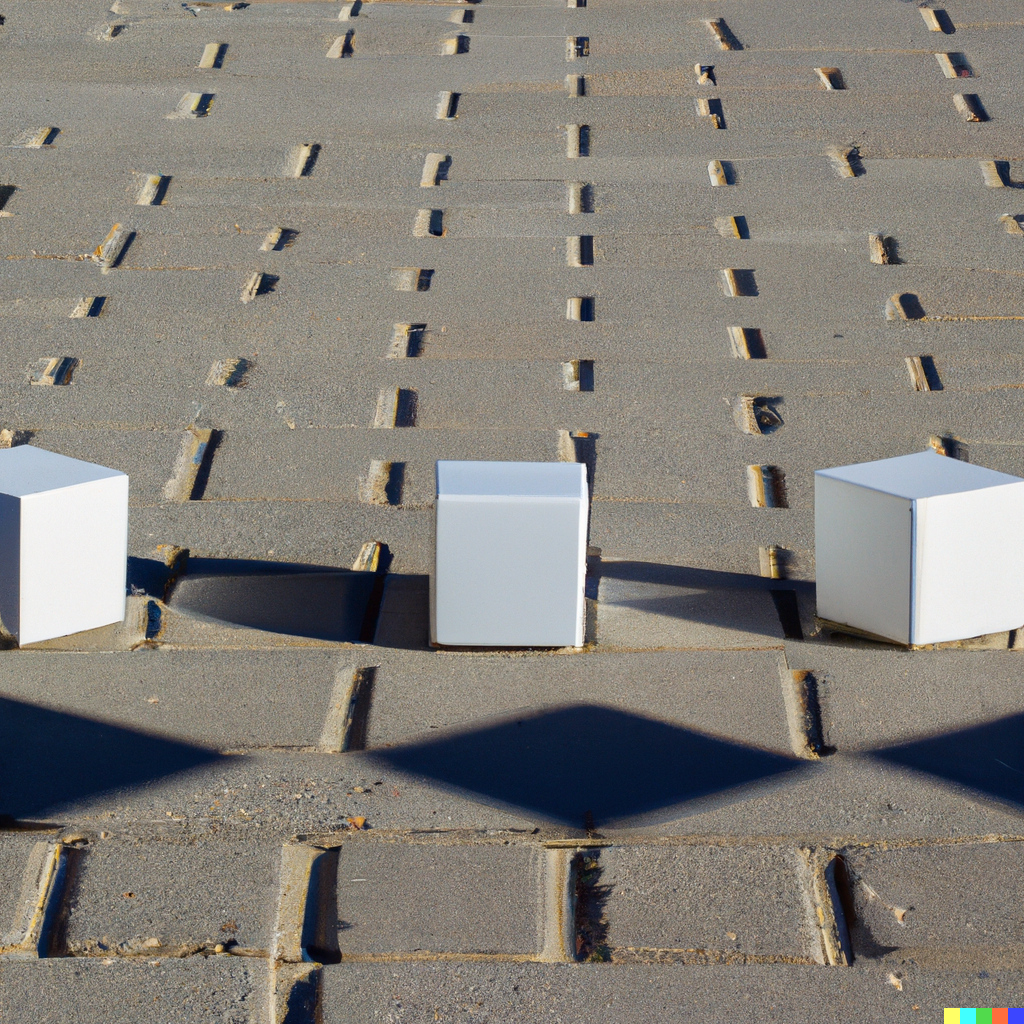} 
    \end{tabular}
    \end{center}
    \caption{Geometrically inconsistent shadows: the shape of the shadows in the top row are inconsistent with the shape of the cubes; the position of the cast shadows in the bottom row (left and center) are shifted away from the cubes; and the shadows in the bottom right are a rare example of a significant failure case.}
    \label{fig:shadows-failure}
\end{figure}

\begin{figure}[t]
    \begin{center}
    \begin{tabular}{ccc}
        \includegraphics[height=9cm]{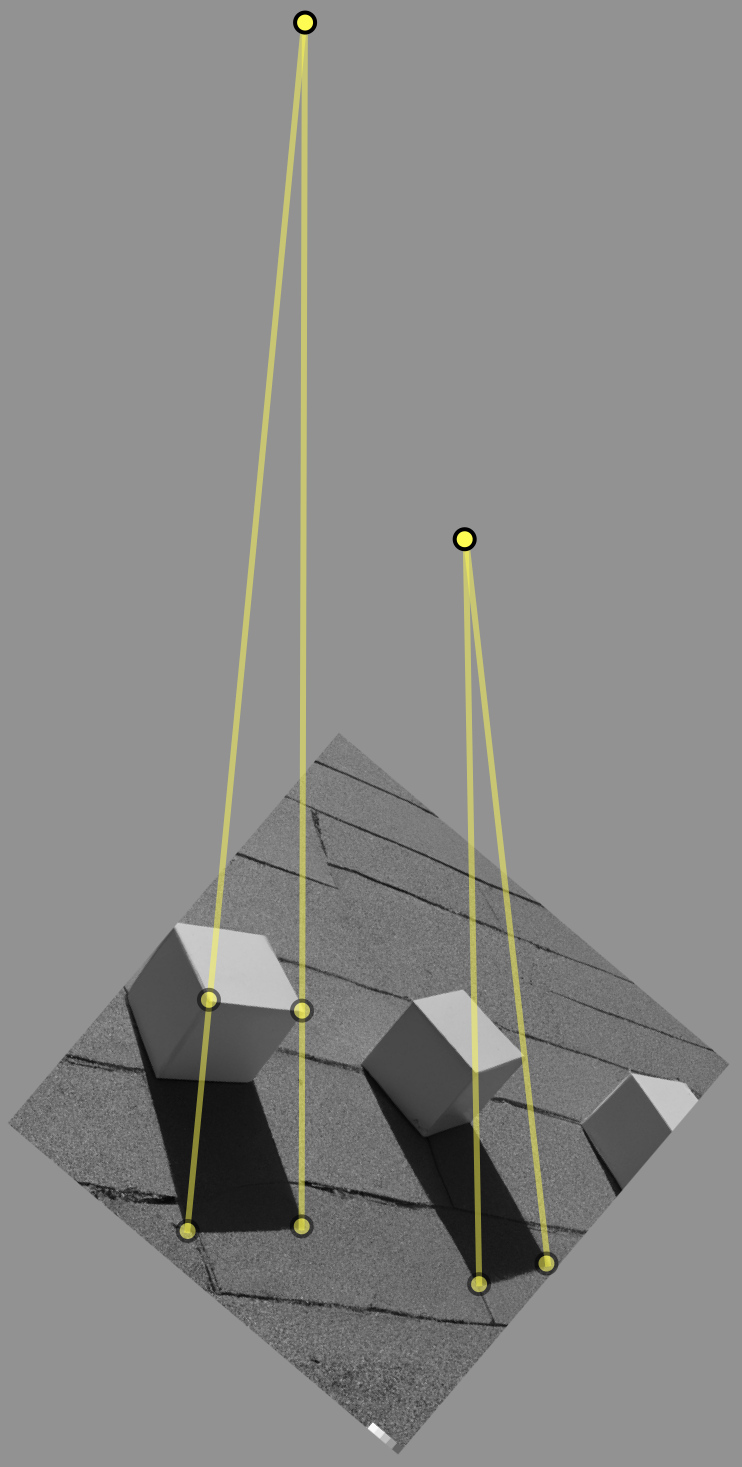} &
        \includegraphics[height=9cm]{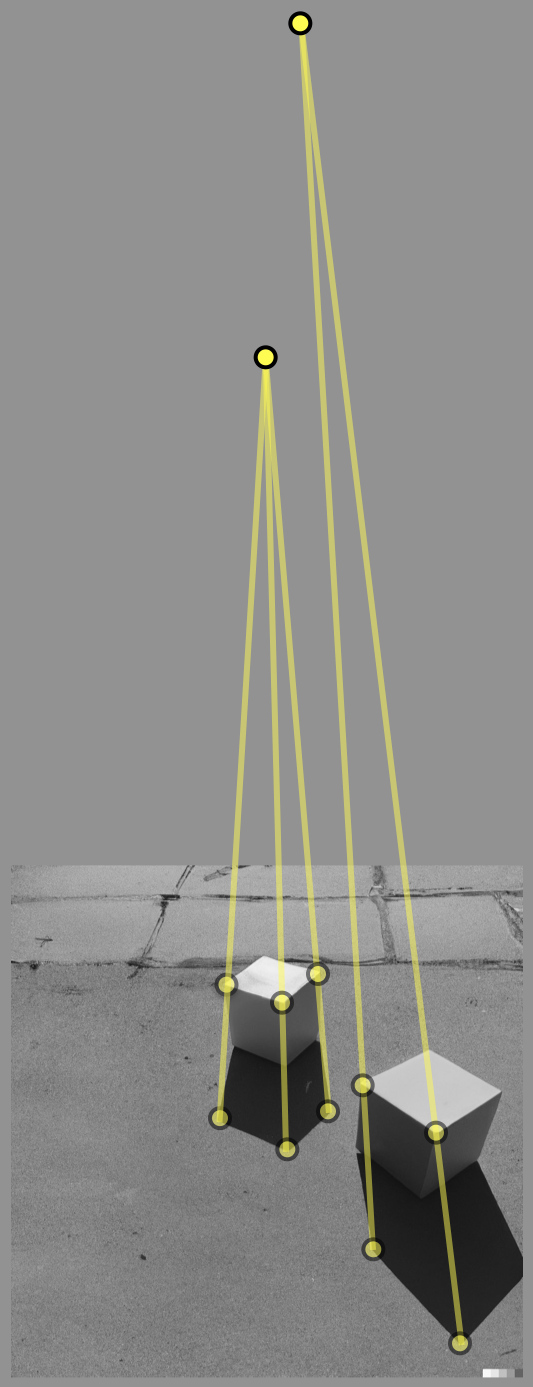} &
        \includegraphics[height=9cm]{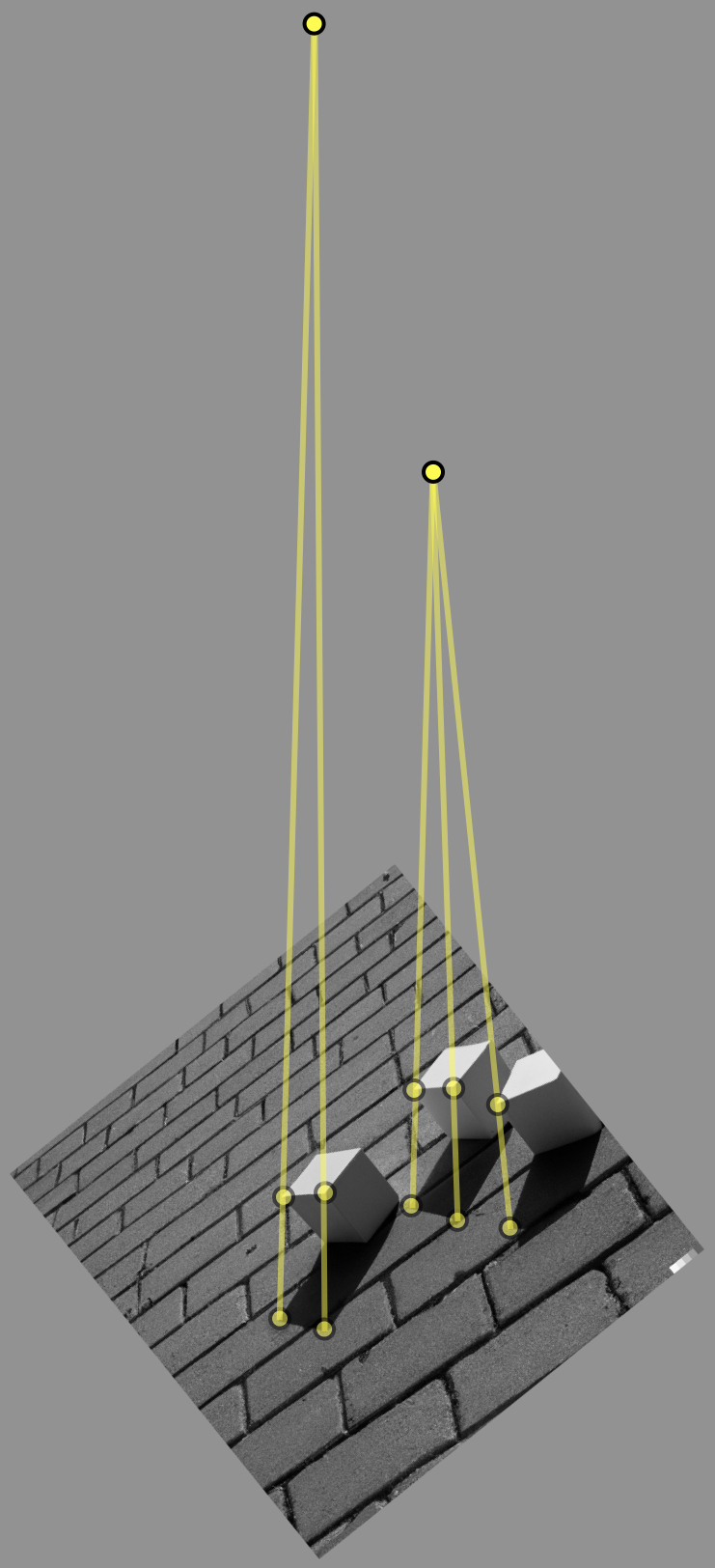} 
    \end{tabular}
    \end{center}
    \caption{Geometrically inconsistent shadows as evidenced by the lack of a consistent intersection of the object-to-shadow constraints. The synthesized images are rotated and converted to grayscale to make the visualizations clearer.}
    \label{fig:shadows-analysis-1}
\end{figure}

A total of $25$ synthesized images with the text prompt ``three cubes on a sidewalk photographed on a sunny day'' were analyzed for cast-shadow consistency. In all cases, the cast shadows are not physically consistent with a scene illuminated by a single light source. Shown in Figure~\ref{fig:shadows-analysis-1} are three representative examples of these shadow inconsistencies in which the cast shadows across objects are visually plausible, but geometrically inconsistent as evidenced by the lack of a single intersection of constraints.

Despite the fact that cast shadows are not physically accurate across the entire scene, it is impressive how perceptually convincing the cast shadows are given the complexity of synthesizing objects with consistent shadow shape and position. There are, nevertheless, some more obvious failure cases. Shown in Figure~\ref{fig:shadows-failure} are representative examples of obviously implausible cast shadows ranging from inconsistent shadow shape to inconsistent shadow position, and a rare (bottom right) example of a significant failure case.

%
%

\subsection{Reflections}
\label{subsec:reflections-analysis}

\begin{figure}[p]
    \begin{center}
    \begin{tabular}{c@{\hspace{0.12cm}}c@{\hspace{0.12cm}}c}
        \includegraphics[width=0.275\linewidth]{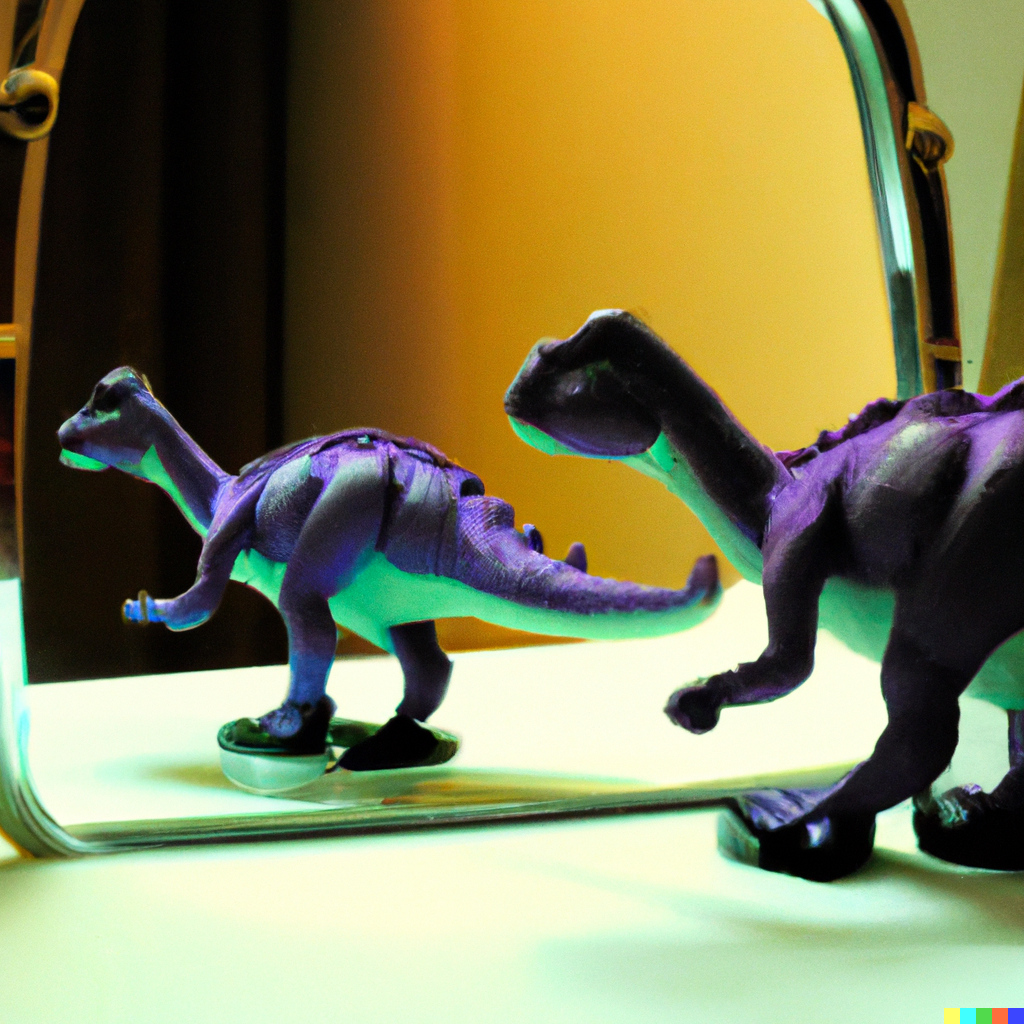} &
        \includegraphics[width=0.275\linewidth]{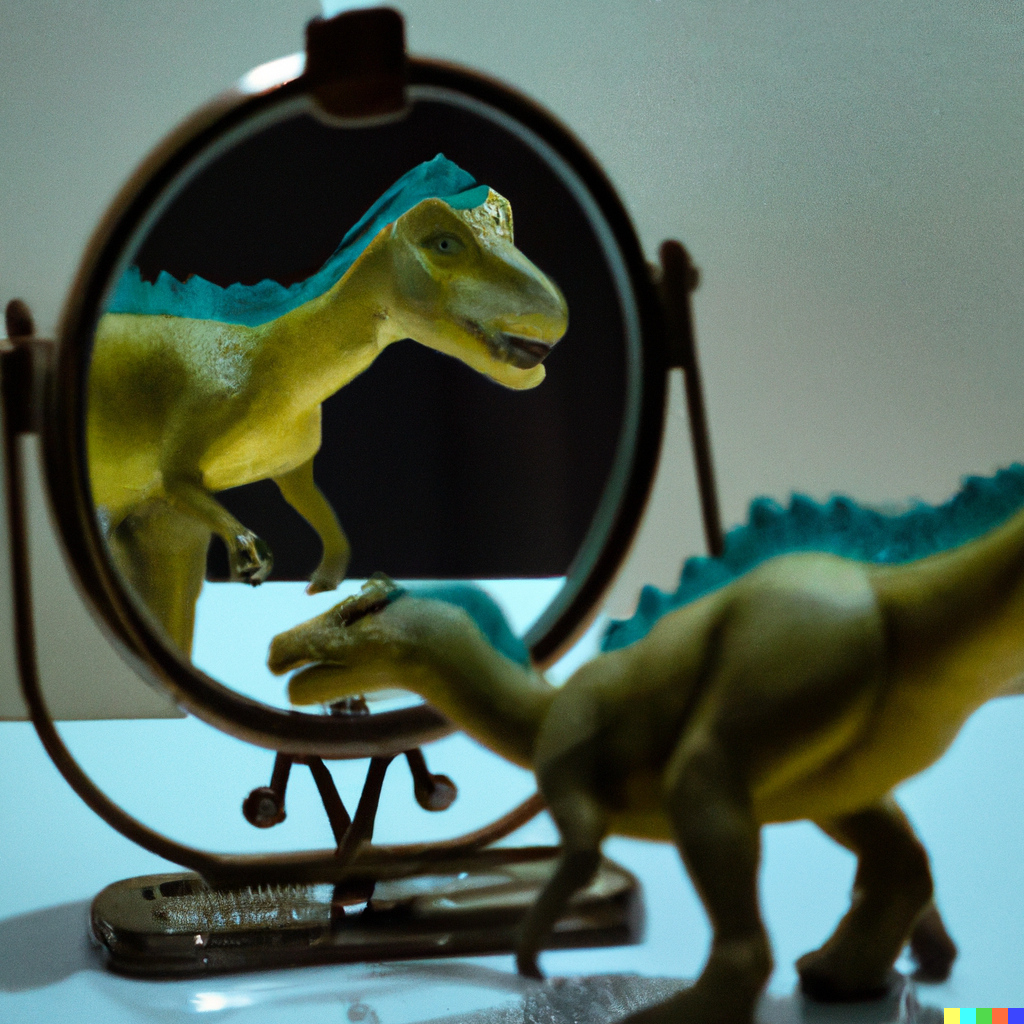} &
        \includegraphics[width=0.275\linewidth]{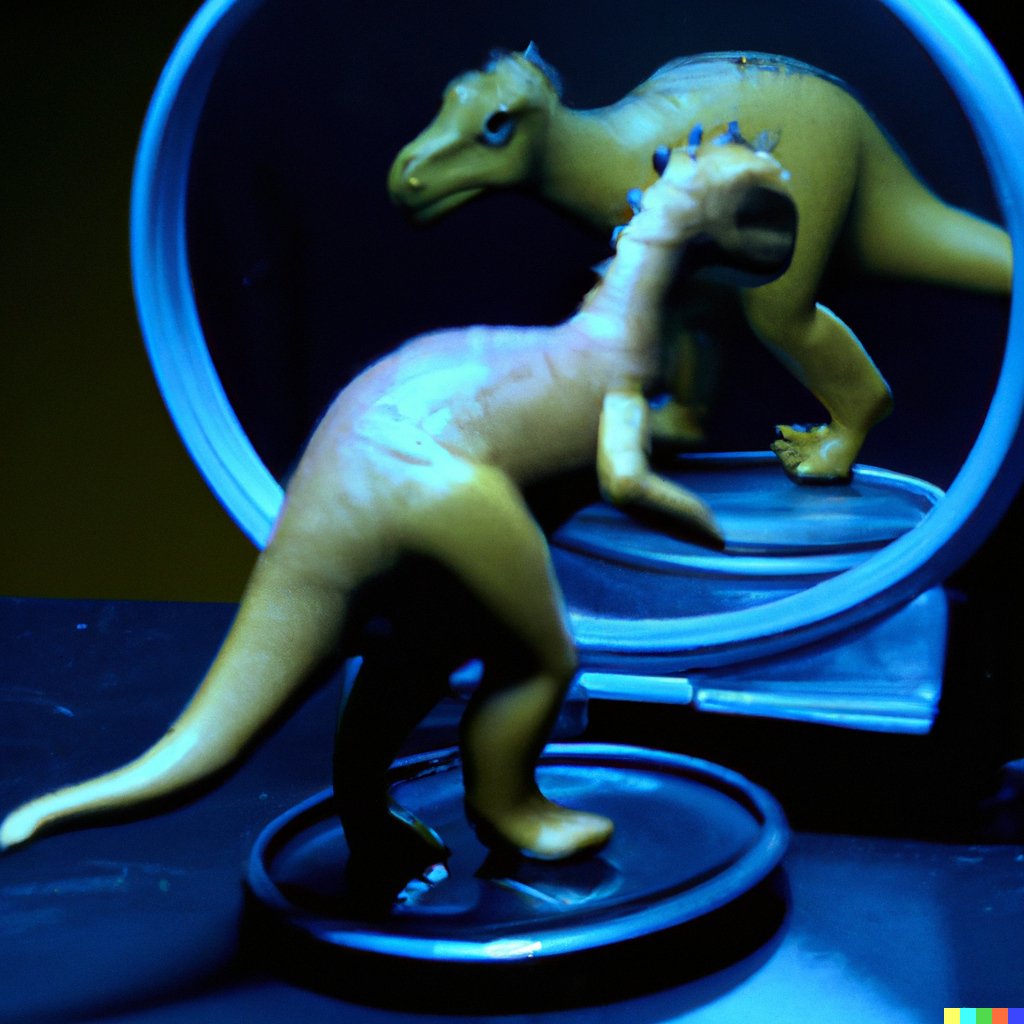} \\
        \includegraphics[width=0.275\linewidth]{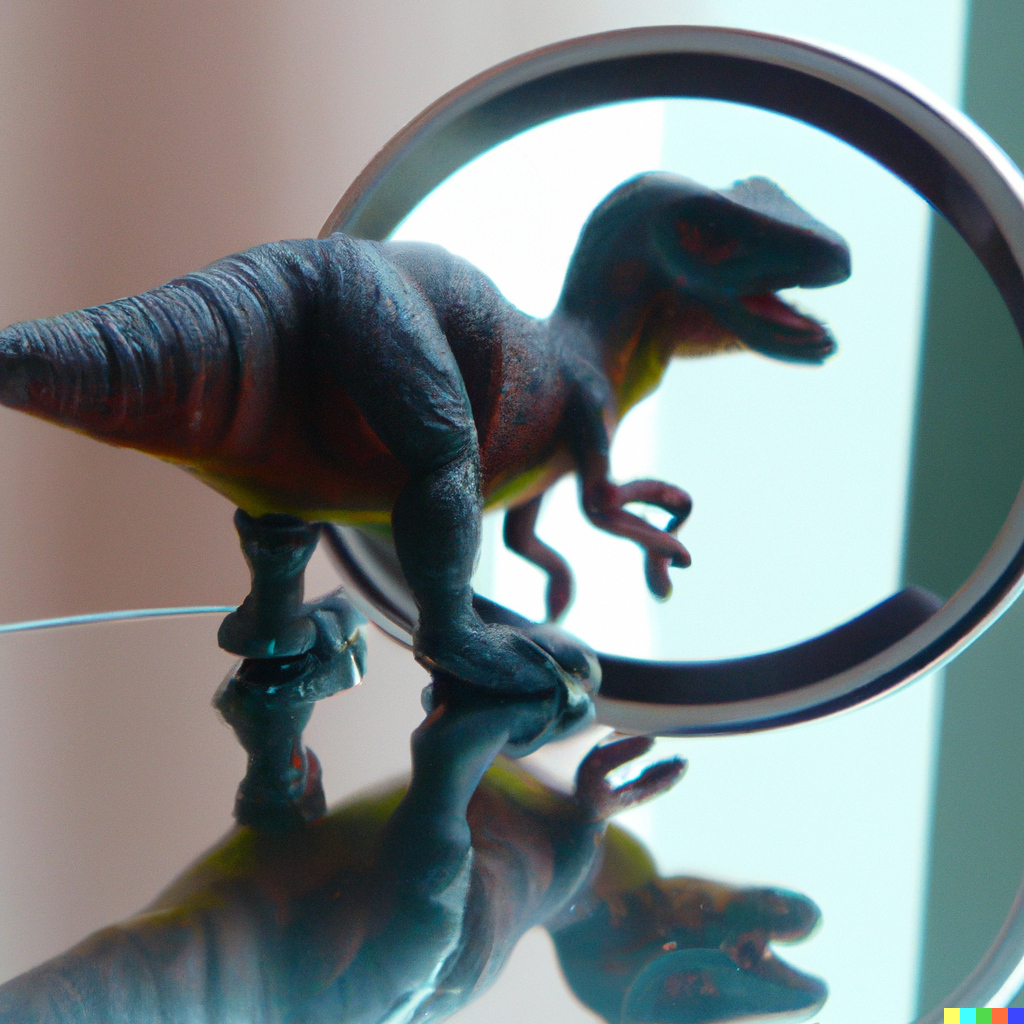} &
        \includegraphics[width=0.275\linewidth]{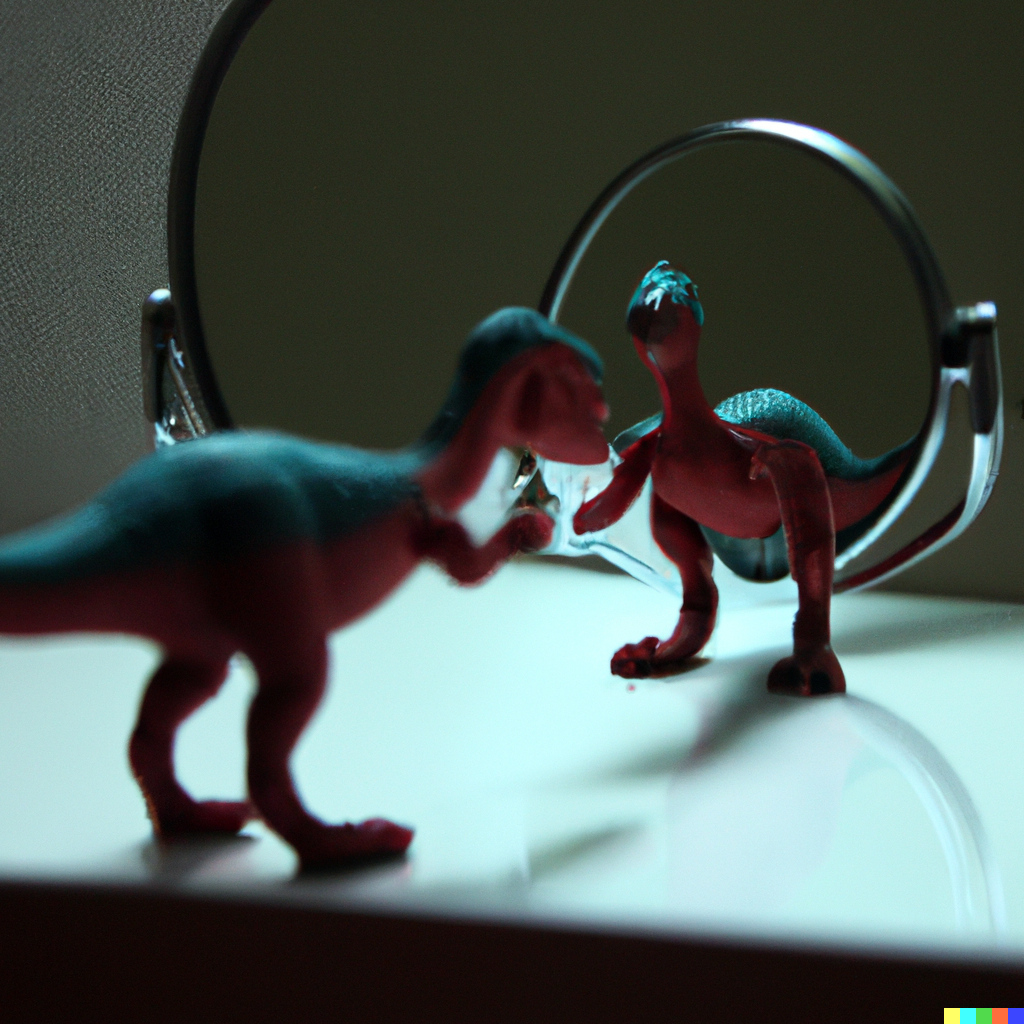} &
        \includegraphics[width=0.275\linewidth]{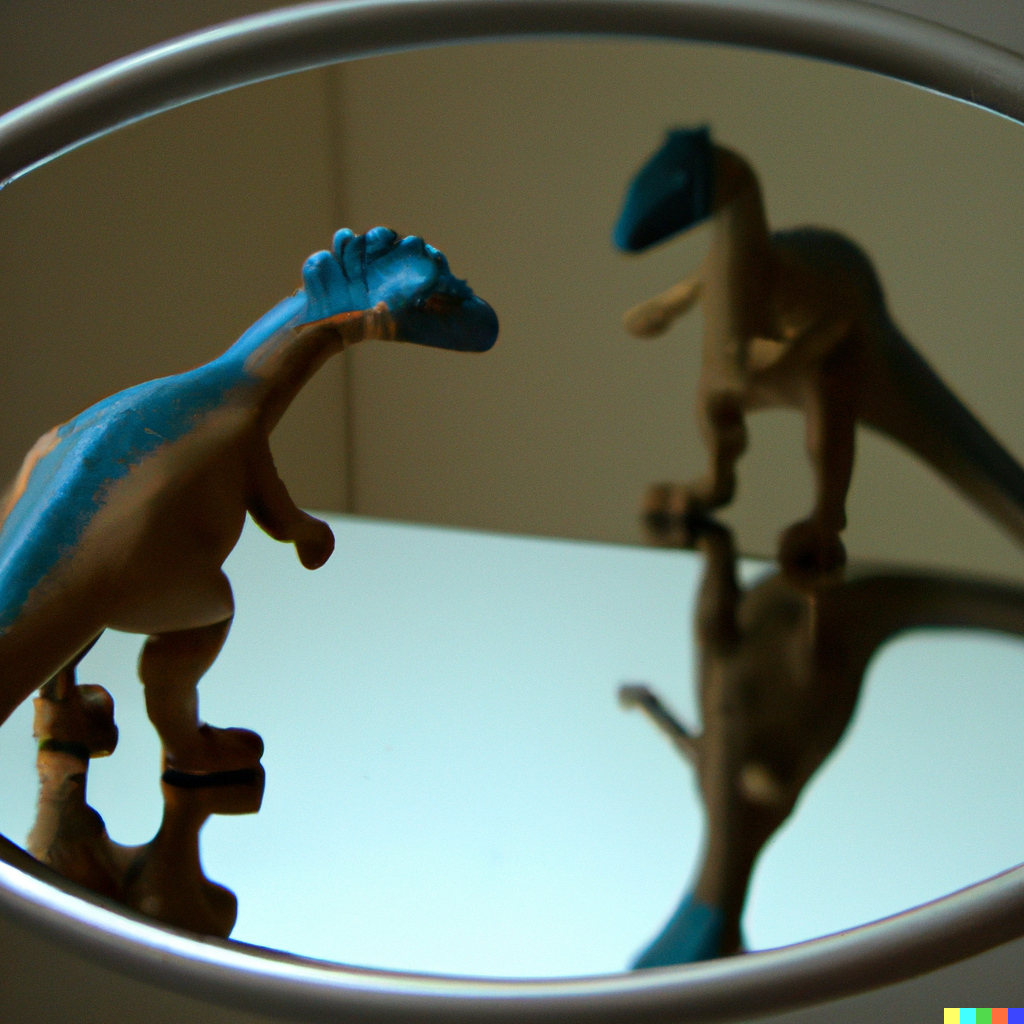} \\
    \end{tabular}
    \end{center}
    \caption{Geometrically implausible reflections.}
    \label{fig:reflections-failure}
\end{figure}

A total of $25$ images were synthesized with the text prompt ``a photo of a toy dinosaur and its reflection in a vanity mirror.'' In most cases, the resulting image contained a toy dinosaur and a mirror, but in all but three cases, the reflection was obviously disconnected from the scene. Shown in Figure~\ref{fig:reflections-failure} are representative examples of these failures. Different text prompts requesting a range of objects and mirror configurations yielded similar results.

\begin{figure}[p]
    \begin{center}
    \begin{tabular}{c}
        \includegraphics[width=0.85\linewidth]{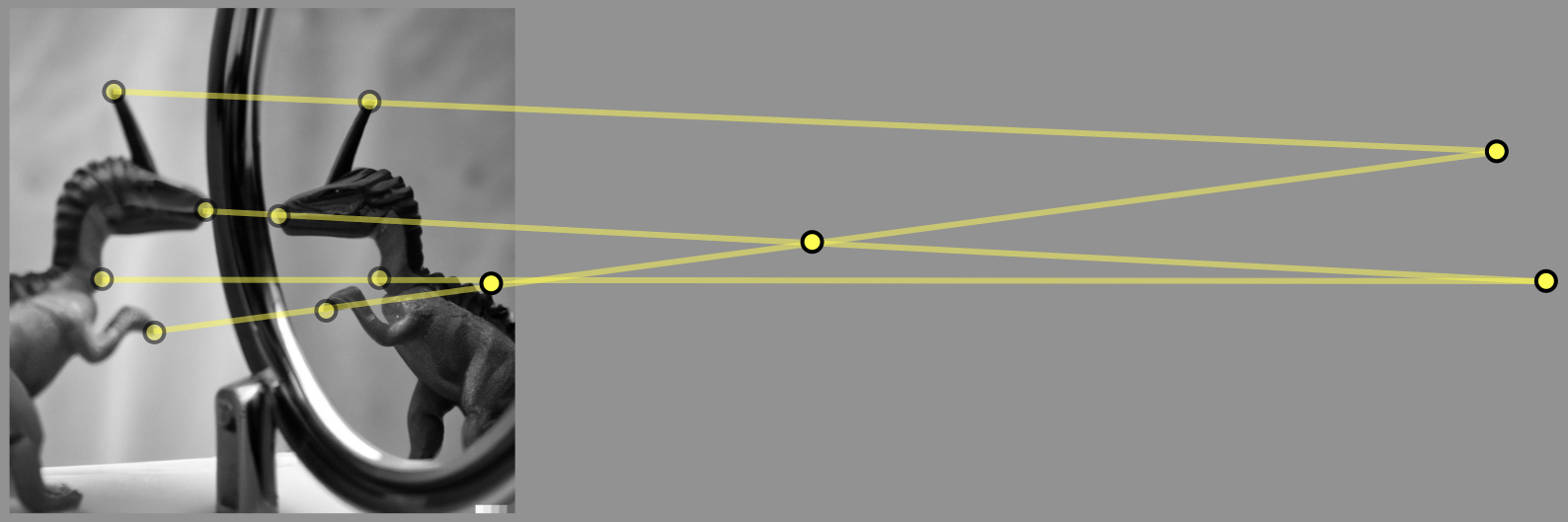} \\
        \includegraphics[width=0.85\linewidth]{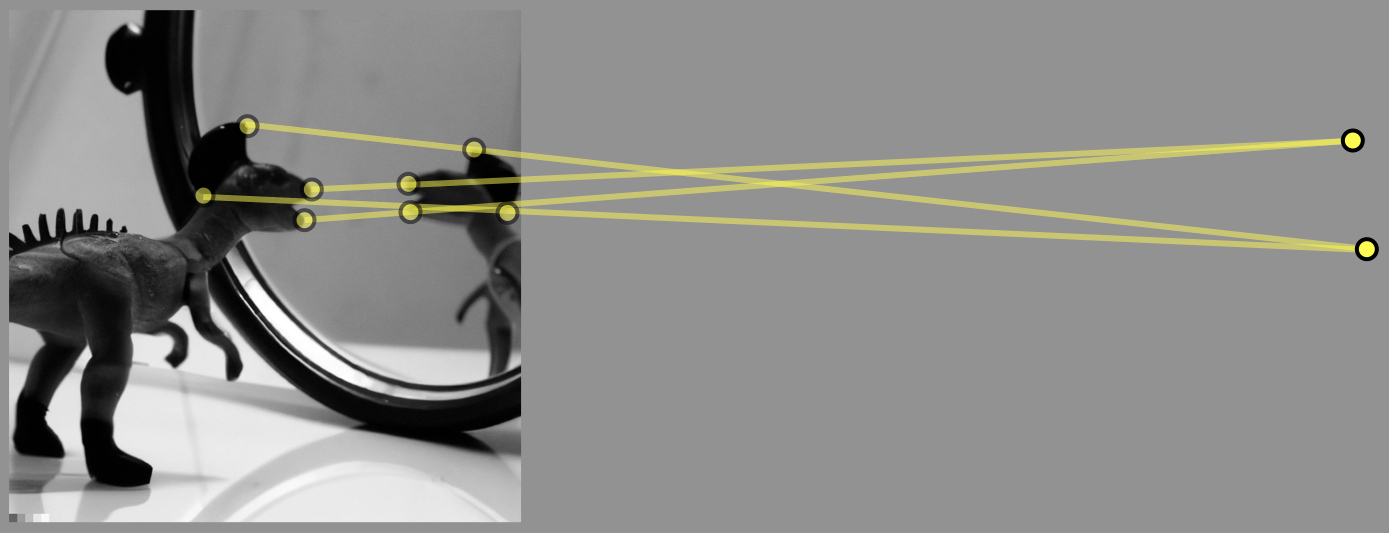} 
    \end{tabular}
    \end{center}
    \caption{Geometrically inconsistent reflections as evidenced by the lack of a consistent intersection of the object-to-reflection constraints. The synthesized images are converted to grayscale to make the visualizations clearer.}
    \label{fig:reflections-analysis}
\end{figure}

Shown in Figure~\ref{fig:reflections-analysis} are the results of applying a geometric analysis to a pair of images that contained reasonably consistent reflections. Although these reflections are visually plausible, the reflections are not geometrically consistent. Unlike the cast shadows and geometric structures in the previous sections, \dalle struggles to synthesize plausible reflections, presumably because such reflections are less common in its training image data set.

%
%

\section{Discussion}
\label{sec:discussion}

The visual relevance and realism of images synthesized by \dalle, and similar paint-by-text techniques, is remarkable. Even in the absence of explicit 3-D modeling of objects, surfaces, and lighting -- as found in traditional CGI-rendering -- synthesized photos exhibit many of the properties of natural scenes. At the same time, geometric structures, cast shadows, and reflections in mirrored surfaces are not fully consistent with the expected perspective geometry of natural scenes. Geometric structures and shadows are, in general, locally consistent, but globally inconsistent. Reflections, on the other hand, are often rendered implausibly, presumably because they are less common in the training image data set. 

The trend in paint by text has been that increasingly larger synthesis engines (defined by the number of model parameters) yields increasingly more realistic images. Increasing the number of model parameters from $350$ million to $20$ billion in Google's Parti, for example, leads to significant improvements in photo realism. Given this trend, it may just be a matter of time before paint-by-text synthesis engines learn to render images with full-blown perspective consistency. Until that time, however, geometric forensic analyses may prove useful in analyzing these images.

On the other hand, given the human visual system's relative insensitivity to perspective inconsistency, paint-by-text synthesis engines could explicitly encode slight perspective inconsistencies in their photo-realistic images to make downstream forensic identification easier.

\section*{Acknowledgement}

Thanks to OpenAI for providing access to the \dalle synthesis engine.

\bibliographystyle{unsrt}
\bibliography{refs}

\end{document}